\documentclass[12pt,preprint,prd,showpacs,showkeys]{revtex4}
\usepackage{amsmath}
\usepackage{graphicx}

\begin{document}
\title{Constraints on  Models for TeV Gamma Rays from Gamma-Ray Bursts}
\author{P. Chris Fragile}
\email{fragile@llnl.gov}
\affiliation{University of California,
Lawrence Livermore National Laboratory, Livermore, CA 94550}
\author{Grant J. Mathews and John Poirier}
\affiliation{University of Notre Dame, Center for Astrophysics,
Department of Physics, Notre Dame, IN 46556}
\author{Tomonori Totani}
\affiliation{Theory Division, National Astronomical Observatory, Mitaka, Tokyo, 181-8588, Japan}

\begin{abstract}
We explore several models which might be proposed to explain recent 
possible detections of
high-energy (TeV) gamma rays
in association with low-energy gamma-ray bursts (GRBs).
Likely values (and/or upper limits) for the source energies
in low- and high-energy gamma rays and hadrons are deduced 
for the burst sources associated with possible TeV 
gamma-ray detections
by the Project {\it GRAND} array.  Possible spectra for 
energetic gammas are deduced
for three models: 1) inverse-Compton scattering of ambient photons from 
relativistic electrons; 2) proton-synchrotron 
emission; and 3) inelastic scattering of relativistic protons from 
ambient photons creating high-energy neutral pions, which decay 
into high-energy photons.  These models rely on some basic assumptions 
about the GRB properties, e.g. that:  the low- and high-energy 
gamma rays are produced at the same location; the time variability 
of the high-energy component can be estimated from the FWHM of 
the highest peak in the low-energy gamma ray light curve; and the 
variability-luminosity relation of Fenimore \& Ramirez-Ruiz (2000) 
gives a reliable estimate of the redshifts of these bursts.
We also explore the impact of each of these assumptions 
upon our models.  
We conclude that the energetic requirements are difficult to satisfy 
for any of these models unless, perhaps, either the photon beaming angle is 
much narrower for the high-energy component than for the low-energy GRB or 
the bursts occur at very low redshifts ($\lesssim 0.01$).  
Nevertheless, we find that the energetic requirements are 
most easily satisfied
if TeV gamma rays are produced predominantly by
inverse-Compton scattering with a magnetic field strength well 
below equipartition or by 
proton-synchrotron emission with a magnetic field strength near 
equipartition.  
\end{abstract}

\maketitle

\section{Introduction}
\label{sec:level1}

Evidence has been accumulating for the arrival
of $\sim$ GeV-TeV gamma rays in coincidence with low-energy
($\sim $ MeV) gamma-ray bursts (GRBs).
For example, {\it EGRET} detected
seven GRBs which emitted high energy photons in the
$\sim 100$ MeV to 18 GeV range
\citep{Schneid,  Hurley, Catelli}.
There have also been results suggestive of
gamma rays beyond the GeV range \citep{Amenomori,
Padilla}, although these were not 
claimed as firm detections.
Evidence  has also been reported for TeV emission in one burst
out of 54 {\it BATSE} GRBs in the field of view of the {\it Milagrito}
detector \citep{milagro}.  In another paper \citet{Poirier}
reported suggestive evidence for sub-TeV gamma rays arriving
in coincidence with GRBs which occurred near zenith above the
Gamma Ray Astrophysics at Notre Dame ({\it GRAND}) air shower
array.   In that experiment, most of the eight bursts analyzed were associated 
with at least some marginal excess ($\sim  1\sigma$) of muons including
the event detected by {\it Milagrito}.  One burst 
evidenced a possible detection at the $2.7\sigma$ level.
As shown in \citet{Poirier},  if this detection is real, then the
output in energetic gammas is likely to dominate the energetics of the burst.

Although these data are not overwhelmingly convincing,
they are at least suggestive that detectable energetic TeV gamma rays might
be associated with low-energy gamma-ray bursts \citep{Vernetto}.
Moreover, these new detections, if real,
can not be explained by a simple extrapolation
of the {\it BATSE} spectrum \citep{Poirier},
particularly if intergalactic absorption
is taken into account \citep{Salamon, tot00}.
A new $\sim$ TeV component in the GRB spectrum seems to be required.

The present work is therefore  an attempt to interpret these new
possible detections in the context of  three models, 
which might be proposed 
for the production of TeV gammas in a GRB.  These are:
1) inverse-Compton scattering of ambient photons from relativistic 
electrons in the burst environment;
2) proton-synchrotron emission \citep{Vietri97, tot98a, tot98b, tot00}; and 3) 
inelastic scattering of relativistic protons from ambient photons 
creating high-energy neutral pions, which decay into high-energy 
photons \citep{Waxman95, Waxman97}.  
We here briefly outline the underlying physics and
characteristic energetic gamma-ray spectra associated with each of these
possible models.  We then derive limits on the parameters of these models
based upon the detection limits from the Project {\it GRAND} array.  Based upon this,
we conclude that it is difficult for any of these models to satisfy 
the energetic requirements unless the photon beaming angle is very 
narrow for the high-energy component.
Of the models considered, the most likely are inverse-Compton scattering or
proton-synchrotron emission.  We note, however, that these conclusions 
rely upon a few assumptions.  For example, we have assumed that 
the low- and high-energy 
gamma rays are produced at the same location; that the time variability 
of the high-energy component can be estimated from the FWHM of 
the highest peak in the low-energy gamma ray light curve; and that the 
variability-luminosity relation of Fenimore \& Ramirez-Ruiz (2000) 
gives a reliable estimate of the redshifts of these bursts.  We 
explore the impact of each of these assumptions and find that, unless 
the bursts occur at very low redshifts 
($\lesssim 0.01$), the 
energetic requirements remain difficult to satisfy.

\section{Low-Energy  GRB Properties}

The mystery of the astrophysical origin for low-energy  gamma ray bursts (GRBs)
has been with us  for some time.
As of yet there is no consensus explanation, although there is mounting
evidence for an association with supernovae \citep{Garnavich02}.
A likely scenario is a burst environment involving
collisions  of an ultra relativistic $e^+-e^-$ plasma fireball
\citep{Paczynski, Goodman, Sari}.
These fireballs may produce low-energy gamma rays either by ``internal''
collisions of multiple shocks \citep{Paczynski94, Meszaros},
or by ``external'' collisions of a single shock with ambient
interstellar material \citep{Rees}.

In either of these paradigms it is possible for very energetic gammas to be
produced through inverse-Compton scattering of ambient photons off the
relativistic electrons.  Furthermore, it seems
likely that baryons would be accelerated along with
the pair plasma to very high energies \citep{Vietri95, Waxman95, tot98a}.
Synchrotron emission from energetic protons
\citep{Vietri97, bottcher, tot98a, tot98b} or possibly 
hadro-production of pions in the burst environment \citep{Waxman97}
and subsequent $\pi^0$ gamma decay 
could also  
lead to an additional spectral component of very
energetic gammas.  In any case,
it is at least plausible that energetic gammas
could arrive in coincidence with a gamma-ray burst.  This was 
the premise of Project {\it GRAND}'s search for high-energy gammas 
arriving in coincidence with {\it BATSE} GRB observations.

It is also possible, however, that low- and high-energy gamma-ray 
components are generated in different regions or phases of a burst.  
This could lead to substantially different arrival times for each 
component.  This was in fact the case for the 18 GeV photon 
observed by {\it EGRET}, which arrived $\sim 4500$ s after the low-energy 
emission had ended.

Observations of TeV gammas could provide important 
clues as to the baryon loading,
Lorentz factor, and ambient magnetic field of the relativistic fireball.
Our goal in this paper is to constrain the possible 
spectrum and source of energetic photons using the Project {\it GRAND} data.  
Hence, we restrict ourselves to considering high-energy gammas 
produced concurrently with low-energy gammas 
consistent with the search technique employed by 
Project {\it GRAND}.

\section{Fits to Observed GRB Spectra}


Table \ref{BATSETab} summarizes some of the features of
the {\it BATSE} and Project {\it GRAND} observations of
the eight GRBs analyzed in \citet{Poirier}, where a  
detailed explanation of the Project 
{\it GRAND} results can be found.
In the present paper, we  will
denote quantities in the frame of
the observer by the superscript ``$ob$''.  We will
also use $\epsilon_\gamma$ to denote the low energy GRB photons and
distinguish them from the high-energy component, denoted $E_\gamma$.  The
observed
{\it BATSE} spectra are fit with
a broken power law of the form \citep{band}
\begin{equation}
\frac{d \phi_\gamma (\epsilon^{ob}_\gamma)}{d
\epsilon^{ob}_\gamma} = a
\begin{cases}
\left( \epsilon^{ob,{\rm MeV}}_{\gamma b}
\right)^{\beta_l-\beta_h} \left( \epsilon^{ob,{\rm MeV}}_\gamma \right)^{-\beta_l}~, & \text{if $\epsilon^{ob}_\gamma <
\epsilon^{ob}_{\gamma b}$,} \\
\left( \epsilon^{ob,{\rm MeV}}_\gamma \right)^{-\beta_h}~, &
\text{if $\epsilon^{ob}_\gamma \ge \epsilon^{ob}_{\gamma b}$,}
\end{cases}
\label{batsespec}
\end{equation}
where $\beta_h \approx 2$, $\beta_l \approx 1$, and
$\epsilon^{ob}_{\gamma b} \approx 1$ MeV is the break energy of
the observed spectrum.  Although these bursts are often better fit by using
an exponential to join the two components, a broken power law is adequate
for the present
discussion.  It maintains a simple analytic form for the equations, and
as we shall see, the precise low energy form is almost irrelevant as
long as a break energy exists.
In what follows
we will use values of $a$, $\beta_h$, $\beta_l$, and $\epsilon_{\gamma b}^{ob}$
  derived from optimum fits to the {\it BATSE} spectra for all events except the
{\it Milagrito} event for which the {\it BATSE} fluence
was too weak to obtain a reliable spectral fit.
The {\it BATSE} fit parameters corresponding to equation (\ref{batsespec})
are listed in Table \ref{gammaraytab}.  The fit parameters for these
bursts were provided at our request by M. S. Briggs at the
Marshall Space Flight Center.  We also include the variability
time scale $\Delta t$ for these bursts, which was estimated as the full width
at half maximum (FWHM) of the brightest peak in each light curve.  
The light curves of GRBs typically show a wide range of 
timescale variability, so this choice may not be justified.  In 
\S \ref{sec:results} we explore the dependence of our 
conclusions on a wide range of values for $\Delta t$.

In Table \ref{zTab}, we give estimated redshifts for each 
burst.  GRB 990123 is the only burst
in this group for which an optical counterpart was detected.  
This burst, therefore, is known
to have  occurred at a redshift of $z=1.6$ \citep{Kulkarni}.  The
redshifts for most of the remaining bursts were estimated using the 
variability-luminosity relation of \citet{Fenimore}.  This
method relies on the apparent correlation between the time
variability of a burst, which can be measured from its light
curve, and the absolute luminosity of the burst, which we wish to
infer. This provides us with
a straightforward method for converting GRB observables into
luminosities and redshifts \citep{Fenimore}.

Following \citet{Fenimore}, 
we first fit a quadratic polynomial to the 
background in the non-burst portions of the {\it BATSE} 64 ms four channel 
data (i.e. DISSC data).  Let $b_i$ be the binned background counts from this 
polynomial fit.  If $g_i$ are the observed binned counts during the actual
burst event, then the net count is $c_i = g_i - b_i$.  We then rebin the 
counts by dilating the time samples by $Y=(1+z)/(1+z_b)$, where $z$ is the
redshift we wish to estimate and $z_b$ is a baseline redshift.  Following
\citep{Fenimore}, we take $z_b=2$.  The new, dilated net 
counts, $C_i$, represent what the time history would look like at $z=z_b$.  
The variability is then defined to be the average mean-square of the 
variations in $C_i$ relative to a smoothed time history, as
\begin{equation}
V = Y^{-0.24}\frac{1}{N} \sum \frac{(C_i-<C>)^2-B_i}{C_p^2} ~,
\label{veq}
\end{equation}
where $C_p$ is the peak of the dilated net count during the burst and
$<C>$ is the count smoothed with a square-wave window with a length equal to
15\% of the duration of the burst.  The $Y^{-0.24}$ term corrects the
variability for the energy-dependence of the time scale of a GRB 
\citep{Fenimore}.  The $B_i$ term (dilated background 
counts in a sample) accounts for the Poisson noise.  The sum is taken
over the $N$ samples that exceed the background by at least $5\sigma$.
The estimated variabilities are listed in Table \ref{zTab}.

Based upon the fits of \citet{Fenimore}, we can relate this
variability $V$ to the peak isotropic luminosity $L_{256}$ 
averaged over 256 ms
in a specified energy range, $\epsilon_{l,p}$ to $\epsilon_{u,p}$ 
(i.e. $L_{256}$ erg s$^{-1}$ in the 50 to 300 keV band).  This 
variability-luminosity relation is
\begin{equation}
L_{256}/(4\pi) = 1.9 \times 10^{51}V^{0.86} \mathrm{erg~s}^{-1} ~.
\label{L256eq1}
\end{equation}
This peak luminosity depends upon the redshift, 
the observed spectral shape, and the observed peak photon flux $P_{256}$
(also averaged over 256 ms and over the same energy range) as
\begin{equation}
L_{256} = 4 \pi D_z^2 P_{256} <\epsilon_\gamma> ~,
\label{L256eq2}
\end{equation}
where $D_z$ is the co-moving distance and $<\epsilon_\gamma>$ is the 
average photon energy
in the luminosity bandpass per photon in the count bandpass.  
For this work we compute luminosities for an
isotropic burst environment, such that $\Omega=4\pi$, where $\Omega$
is the unknown opening angle of the burst.  
If GRBs emit in a jet, our inferred luminosities 
and energies are diminished by $\Omega/4\pi$.

The co-moving distance $D_z$ 
for a flat $\Omega_M +
\Omega_\Lambda = 1$ model is simply given by
\begin{equation}
D_z = \frac{c}{H_0} \int^z_0 \left[\Omega_M (1+z')^3
+\Omega_\Lambda \right]^{-1/2} dz'~.
\label{dzeq}
\end{equation}
For the purposes of the present discussion, we will adopt the
currently popular $\Omega_M =0.3$, $\Omega_\Lambda =0.7$, $H_0
=72$ km s$^{-1}$ Mpc$^{-1}$ model \citep{Garnavich, Perlmutter, 
Freedman}.  From the observed photon spectrum the average photon 
energy in the luminosity bandpass per photon in the count bandpass is 
\begin{equation}
<\epsilon_\gamma> = \frac{\int_{\epsilon_{l,p}}^{\epsilon_{u,p}} 
\epsilon_\gamma \phi[\epsilon_\gamma/Y] d\epsilon_\gamma}
{\int_{\epsilon_l}^{\epsilon_u} \phi[\epsilon_\gamma] d\epsilon_\gamma} ~,
\label{aveeq}
\end{equation}
where $\epsilon_l$ and $\epsilon_u$ are the limits on the {\it BATSE} energy
range ($\approx 20-1500$ keV).
We can now iteratively solve equations (\ref{veq}-\ref{aveeq}) until the 
estimate for $z$ converges.

This method of estimating the redshift of GRBs was found 
\citep{Reichart} to be
consistent with other estimates that rely upon an apparent
relation between the luminosity and the time lag between hard and
soft energy peaks.  The correlation between
these two independent methods argues in favor of their reliability
\citep{Schaefer}.  However, in \S \ref{sec:results} 
we explore the impact of systematically larger 
and smaller redshifts on our conclusions.

Additionally, we can estimate the effective $4\pi$ luminosity
at the source in the {\it BATSE} energy band time-averaged over the full $T90$ 
interval, $L_\gamma$, by
\begin{equation}
L_\gamma = 4\pi D^2_z \int_{\epsilon_l}^{\epsilon_u} 
d\epsilon_\gamma \epsilon_\gamma
\frac{d\phi[\epsilon_\gamma/(1+z)]}{d\epsilon_\gamma}~. \label{lumeq}
\end{equation}
This is the luminosity estimate we use for the rest of the work presented
in this paper.  Both luminosity estimates are listed in Table \ref{zTab}
for each burst.

\subsection{{GRB 970417a}}
GRB 970417a was the one burst (of 54 in the field of view) for which the 
{\it Milagrito} collaboration reported evidence of TeV emission during the 
duration of this burst within the {\it BATSE} error circle 
\citep{milagro}.
  For this reason,
we have included it in our analysis.  Interestingly, this was a
relatively weak {\it BATSE} burst with a fluence of $3.9 \times 10^{-7}$
erg cm$^{-2}$ in all four {\it BATSE} energy channels ($>20$ keV).  Because
this is such a weak low-energy burst, it is difficult to obtain reliable
fits for the {\it BATSE} spectral parameters.  For this reason, we have
instead used the following average values of all bright bursts 
from \citet{Preece}:
$\beta_l = 1.0$, $\beta_h = 2.25$, and $\epsilon^{ob}_{\gamma b} = 225$ keV.
We can then use these average parameters and the observed flux to fix the
normalization, $a$.  The weak signal in the {\it BATSE} band also 
prevents
us from using the variability-luminosity relation described above to 
determine the
redshift of this burst.  Instead we adopt
$z=0.7$ based upon the analysis of \citet{tot00}.

\section{The Models}

\subsection{Inverse-Compton Spectrum}
One possible source for energetic gamma rays is
the inverse-Compton (IC) scattering of low-energy ambient photons
by relativistic electrons.  Indeed, such IC photons are
thought to be the source of observed high-energy photons from
active galactic nuclei such as Mk-421 \citep{Zdziarski}.

The inverse-Compton-scattering spectrum is generally written as
\begin{equation}
\left(\frac{d \Phi_\gamma}{d E_\gamma}\right)_{\rm IC} \propto 
E_\gamma^{-1} \int dE_e \frac{d \Phi_e}{d E_e} \int dx f_{e-s}(x) ~~,
\end{equation}
where $d \Phi_e/dE_e$ is the spectrum of electrons characteristic of 
a Fermi mechanism, 
\begin{equation}
\frac{d \Phi}{dE} = A
\begin{cases}
E^{-\alpha}~, & \text{if $E < E_b$,} \\
E_b E^{-\alpha - 1}~, &
\text{if $E \ge E_b$,}
\end{cases}
\label{fermi}
\end{equation}
where $A$ is a normalization constant and $\alpha \approx 2 \pm 0.2$
\citep{Hillas}.
The flux of the synchrotron photons $f_{e-s}$ 
is 
\begin{equation}
f_{e-s}(x) = 
\begin{cases}
x^{-(\alpha-1)/2}~, & \text{below the cooling break,} \\
x^{-\alpha/2}~, & \text{above the cooling break,}
\end{cases}
\end{equation}
where $x\propto E_\gamma/E_e^2$.  
Evaluating the integrals, we arrive at the simplified expression  
\begin{equation}
(d \Phi_\gamma/d E_\gamma)_{\rm IC}  \propto E_\gamma^{-\delta}~~,
\end{equation}
where the spectral index $\delta$ is given by
\begin{equation}
\delta =
\begin{cases}
(\alpha+1)/2 ~, & \text{for $E_\gamma < E_{\rm cool}$~,} \\
(\alpha+2)/2~, & \text{for $E_\gamma > E_{\rm cool}$~,}
\end{cases}
\label{icexp}
\end{equation}
   where $\alpha \sim 2$ is the electron spectral index
   and $E_{\rm cool}$ is the energy scale at
which the electron cooling time becomes comparable to the system
lifetime. The inverse-Compton cooling time in the frame of the shock is
\begin{equation}
\frac{1}{t_{\rm IC}} = 1.3 \times 10^{-4} \frac{\gamma_e L_{\gamma,
51}^{ob} (1+z)^4}{\Gamma_{300}^6 \Delta t^2} ~~\mathrm{s}^{-1}~,
\end{equation}
where $\gamma_e$ is the electron Lorentz factor and 
$\Gamma = 300\Gamma_{300}$ is the 
bulk Lorentz factor, both in the shock frame; 
$L^{ob}_{\gamma ,51}=L^{ob}_\gamma /10^{51}$ erg s$^{-1}$ where 
$L^{ob}_\gamma = L_\gamma /(1+z)^2$; and 
$\Delta t = (1+z)r_d/\Gamma^2 c$ is the variability timescale (in seconds) for
the GRB, where $r_d$ is the radius at which low-energy gamma rays are emitted,
both measured in the observer frame.

The total fractional power radiated in inverse-Compton photons 
can be estimated from the ratio of the
expansion time to the cooling time:
\begin{eqnarray}
f_{\rm IC} &=& \frac{\Gamma \Delta t}{(1+z) t_{\rm IC}} \\
&=& 3.9 \times 10^{-2} \frac{\gamma_e L_{\gamma, 51}^{ob}
(1+z)^3}{\Gamma_{300}^{5} \Delta t}~.
\label{fic}
\end{eqnarray}
The cooling frequency corresponds to $f_{\rm IC} = 1$.
The relation between the Compton-scattered photon energy and $\gamma_e$ in
the observer's frame is
$E_\gamma^{ob} \sim \gamma_e^2 \epsilon_\gamma^{ob}$, where again
$\epsilon_\gamma^{ob}$ is the energy of the {\it BATSE} photons in the
observer's frame.  For simplicity we will take $\epsilon^{ob}_\gamma =
\epsilon^{ob}_{\gamma b}$.
Therefore, the cooling break energy can be written as
\begin{eqnarray}
E_{\rm cool}^{ob} = 0.66 \frac{\Gamma_{300}^{10} \Delta t^2
\epsilon_{\gamma b}^{ob,\mathrm{MeV}}}{(L_{\gamma, 51}^{ob})^2 (1+z)^6}
~~\mathrm{GeV}~.
\end{eqnarray}

Relativistic Klein-Nishina (KN) corrections to the Compton spectrum may be
important in the GRB environment over some range of energies.
For completeness we include these by introducing a parameter
$\Gamma_{\rm KN} = E_\gamma^{ob} \epsilon_\gamma^{ob} (1+z)^2 / \Gamma^2
m_e^2 c^4$.
When $\Gamma_{\rm KN} \gg 1$, the KN effect is important.
We adopt $\Gamma_{\rm
KN}=1$ as the the lower boundary of the KN regime.  This implies
a lower limit to the observed gamma energy at which the
KN effects should be considered,
\begin{eqnarray}
E_{\rm KN}^{ob}
\equiv 24 \frac{\Gamma_{300}^{2}}{(1+z)^2 \epsilon_{\gamma b}^{ob,\mathrm{MeV}}}~~
\mathrm{GeV}~.
\end{eqnarray}
In the  KN regime  the emissivity
of IC radiation per electron is independent of the electron energy and the
cooling time becomes proportional to the electron energy
\citep[see e.g.][]{brumen}.
The emissivity is reduced by a factor of $\sim \Gamma_{\rm KN}^{-2}$
compared with the classical IC formula (see Eq. [\ref{fic}]), and
the observed photon energy is $E_\gamma^{ob} \sim \Gamma \gamma_e m_e
c^2/(1+z)$.
We can thus
define the cooling frequency in the KN regime, by setting
$f_{\rm IC}^{\rm KN} = f_{\rm IC} \Gamma_{\rm KN}^{-2} = 1$.
This gives
\begin{equation}
E_{\rm cool}^{{\rm KN},ob} = 140 \frac{L_{\gamma, 51}^{ob}}{\Gamma_{300}^2
\Delta t (\epsilon_{\gamma b}^{ob,\mathrm{MeV}})^2} ~~\mathrm{GeV},
\end{equation}
where we have used $E_\gamma^{ob} \sim \Gamma \gamma_e m_e c^2/(1+z)$ for
the observed Compton-scattered photon energy.

In the KN regime, the cooling is efficient in the lower photon
energy range of  $E_\gamma < E_{\rm cool}^{\rm KN}$,
i.e., contrary to the situation in the non-KN regime.
In the KN regime, the photon index becomes
\begin{equation}
\delta =
\begin{cases}
\alpha ~, & \text{for $E_\gamma < E_{\rm cool}^{\rm KN}$~,} \\
\alpha + 1 ~, & \text{for $E_\gamma > E_{\rm cool}^{\rm KN}$~.}
\end{cases}
\end{equation}

To summarize, the IC spectrum can be modeled as
\begin{equation}
\left( \frac{d\Phi_\gamma}{dE_\gamma} \right)_{\rm IC} = A^{\mathrm{IC}}_\gamma 
\begin{cases}
\left( E^{ob}_\gamma \right)^{-(\alpha+1)/2} ~, & \text{if $E_\gamma < E_{\rm cool}$~,} \\
\left( E_{\rm cool} \right)^{1/2} \left( E^{ob}_\gamma \right)^{-(\alpha+2)/2} ~, & \text{if $E_{\rm cool} \le E_\gamma < E_{\rm KN}$~,} \\
\left( E_{\rm cool} \right)^{1/2} \left( E_{\rm KN} \right)^{\alpha/2 -1} \left( E^{ob}_\gamma \right)^{-\alpha} ~, & \text{if $E_{\rm KN} \le E_\gamma < E^{\rm KN}_{\rm cool}$~,} \\
\left( E_{\rm cool} \right)^{1/2} \left( E_{\rm KN} \right)^{\alpha/2 -1} \left( E^{\rm KN}_{\rm cool} \right) \left( E^{ob}_\gamma \right)^{-(\alpha+1)} ~, & \text{if $E_\gamma \ge E^{\rm KN}_{\rm cool}$~,}
\end{cases}
\label{icspec}
\end{equation}
where all energies are assumed to be measured in GeV.

For typical parameters, all three quantities,
$E_{\rm cool}, E_{\rm KN}$, and $E_{\rm cool}^{\rm KN}$
are around 1-100 GeV.  In any case, the spectrum beyond $E_\gamma \gtrsim$
100 GeV must be steep with $\delta \sim 3$.  
The resultant IC gamma-ray spectrum should cut off above
\begin{equation}
E_{\gamma,{\rm IC}}^{ob,max}  \approx E^{ob,max}_e = 4.2 \times 10^{5} \frac{\Gamma^{5/2}_{300} \Delta t^{1/2}}{\xi^{1/4}_B (L^{ob}_{\gamma,51})^{1/4} (1+z)^2} ~~\mathrm{GeV}~,
\end{equation}
due to the cut-off in the  spectrum of ultra-relativistic electrons at energies above
$ E^{ob,max}_e $.

\subsection{Spectrum of Proton-Synchrotron Gamma Rays}
It is generally believed that the expanding plasma contains at least some
baryons.
Indeed, some baryons within the jet are required to
increase the burst duration and luminosity \citep[cf.][]{Salmonson}.
In the region where the electrons
are accelerated, protons may also be accelerated
up to ultra-high energies $> 10^{20}$ eV \citep{Waxman95} producing
a spectrum characteristic of
a Fermi mechanism, (eq. [\ref{fermi}]).

   The possibility of energetic protons producing $\sim$TeV gammas
by synchrotron emission has been discussed in a number of papers
\citep{Vietri97, bottcher, tot98a, tot98b, tot00}.
This mechanism has the desirable characteristic that
the low-energy photons produced by electron-synchrotron emission and
the high-energy photons from energetic protons can be produced
simultaneously in the same environment \citep{tot98a, tot98b}.

In this model one assumes that there is a magnetic field present
in the burst environment with approximate equipartition between
the magnetic energy density  $B^2/8 \pi$ and
the total energy density, $U$.
  That is,
\begin{equation}
\frac{B^2}{8 \pi} = \xi_B U~,
\label{b2/8pi}
\end{equation}
where $\xi_B$ is a fraction of order unity.  Following \citet{tot98b}
we assume an optimally efficient proton-synchrotron environment in which
$U \sim U_p \sim (m_p/m_e)U_\gamma$, where $U_\gamma$ is the photon energy
density of {\it BATSE} gamma rays in the frame of the burst.  
This latter quantity can be related to the GRB luminosity
utilizing $ L_\gamma = 4 \pi r_d^2 \Gamma^2 c U_\gamma$.  
We can then rewrite Eq. (\ref{b2/8pi}) in terms
of the variables introduced in the previous section.  This gives
\begin{equation}
B=1.4 \times 10^5 \frac{\xi^{1/2}_B (L^{ob}_{\gamma,51})^{1/2} (1+z)^2}{
\Gamma^{3}_{300} \Delta t}~~\mathrm{G}~.
\label{eqb}
\end{equation}
The photon energy from proton-synchrotron emission
in the observer's frame is then
\begin{eqnarray}
E^{ob}_{\gamma,p-s} & = & \frac{\Gamma \Gamma^2_p e \hbar B}{(1+z) m_p c}
\nonumber \\
   & = & \frac{(E^{ob}_p)^2 e \hbar B (1+z)}{m^3_p \Gamma c^5}~,
\label{egammas}
\end{eqnarray}
where $\Gamma_p = E_p/m_pc^2$ is the relativistic gamma factor
of the protons in the frame of the fireball.  
This leads to the desired relation between the observed proton
energy and the observed gamma energy,
\begin{equation}
E^{ob}_p = C \times \left( E^{ob,{\rm GeV}}_{\gamma,p-s}  \right)^{1/2}~,
\label{eps}
\end{equation}
where,
\begin{equation}
C=1.6 \times 10^{9} \frac{\Gamma^2_{300} \Delta t^{1/2}}{\xi^{1/4}_B
(L^{ob}_{\gamma,51})^{1/4} (1+z)^{3/2}}~~\mathrm{GeV}~.
\end{equation}

The final quantity needed to derive the energetic gamma spectrum
is the cooling rate due to synchrotron emission.
In the frame of the shock, the synchrotron cooling rate is
\begin{eqnarray}
\frac{1}{t_\mathrm{p-sync}}  & \equiv & -\frac{1}{E_p} \frac{d E_p}{dt} \nonumber \\
   & = & \frac{4 e^4 B^2 E^{ob}_p (1+z)}{9
m^4_p c^7 \Gamma} \nonumber \\
   & = & 7.5 \times 10^{-2} \frac{\xi^{3/4}_B
   (L^{ob}_{\gamma,51})^{3/4} (1+z)^{7/2}}{\Gamma^5_{300} \Delta
   t^{3/2}} \left( E^{ob,{\rm GeV}}_\gamma \right)^{1/2}~~\mathrm{s}^{-1}~.
\end{eqnarray}
The fractional energy loss to synchrotron photons
then becomes
\begin{eqnarray}
f_\mathrm{p-sync}(E^{ob}_p) & \simeq & \frac{\Gamma \Delta t}{(1+z)
t_\mathrm{p-sync}} \nonumber \\
   & = &
\begin{cases}
\frac{E^{ob}_p}{E^{ob}_{pb,p-s}}~, & \text{if $E^{ob}_p < E^{ob}_{pb,p-s}$,} \\
1~, & \text{if $E^{ob}_p \ge E^{ob}_{pb,p-s}$,}
\end{cases}
\label{fpsynceq}
\end{eqnarray}
where
\begin{equation}
E^{ob}_{pb,p-s} = 7.5 \times 10^{7} \frac{\Gamma^6_{300} \Delta t}{\xi_B
L^{ob}_{\gamma,51} (1+z)^4} ~~\mathrm{GeV} ~.
\end{equation}
Finally, we note that a proton
flux $\Phi_p (E_p)$ will produce an energetic gamma flux of
\begin{equation}
\Phi_\gamma = \frac{E_p}{E_\gamma} f_\pi (E_p)\Phi_p~.
\label{ngamma}
\end{equation}

The observed high-energy photon spectrum can
then be derived from equations (\ref{fermi}), (\ref{ngamma}),
and (\ref{eps}),
\begin{equation}
\frac{d\Phi_\gamma}{dE_\gamma}=\frac{d\Phi_p}{dE_p} \times
\frac{dE_p}{dE_\gamma} \times \frac{d\Phi_\gamma}{d\Phi_p}~,
\label{phigam}
\end{equation}
to yield 
\begin{equation}
\left(\frac{d \Phi_\gamma}{d E_\gamma}\right)_{p-s} = \frac{1}{2}
A^\mathrm{p-sync}_p C^{2-\alpha} \times
\begin{cases}
\left( E^{ob,\mathrm{GeV}}_{\gamma b,p-s} \right)^{-1/2} \left(
E^{ob,\mathrm{GeV}}_\gamma \right)^{-(\alpha+1)/2}~, & \text{if
$E^{ob}_\gamma < E^{ob}_{\gamma b,p-s}$,} \\
\left( E^{ob,\mathrm{GeV}}_\gamma \right)^{-(\alpha+2)/2}~, &
\text{if $E^{ob}_\gamma \ge E^{ob}_{\gamma b,p-s}$,}
\end{cases}
\label{synspec}
\end{equation}
where $A^\mathrm{p-sync}_p$ is a normalization constant to be determined from 
observations.  The break
energy is
\begin{equation}
E^{ob}_{\gamma b,p-s} = 2.1 \times 10^{-3} \frac{\Gamma^8_{300} \Delta
t}{\xi^{3/2}_B (L^{ob}_{\gamma,51})^{3/2} (1+z)^5} ~~\mathrm{GeV} ~.
\end{equation}

The gamma-ray spectrum
should cut off above  $E_{\gamma,p-s}^{ob,max} \approx 15 \Gamma_{300}/(1+z) $ TeV
\citep{tot98b}
due to the cut-off in the  spectrum of ultra-relativistic protons
   at energies above
\begin{equation}
E^{ob,max}_p = 2.1 \times 10^{11} \frac{\Gamma^{5/2}_{300} \Delta
t^{1/2}}{\xi^{1/4}_B (L^{ob}_{\gamma,51})^{1/4} (1+z)^2} ~~\mathrm{GeV}~.
\end{equation}

\subsection{Spectrum of Photo-Pion Gamma Rays}

Along with undergoing synchrotron radiation, protons accelerated in 
the burst environment may collide with photons in the expanding
fireball to produce secondary pions,
which subsequently decay into
high-energy gammas and neutrinos.  This source of gamma rays 
seems unlikely due to 
the fact that it results from a secondary strong interaction and 
therefore has a small cross-section relative to electromagnetic 
interactions.  Nevertheless an estimate of this spectrum is 
straightforward, so we include it here.
Another alternative possibility might be pion production via
proton-proton collisions \citep[cf.][]{Paczynski94}.  However, the proton
density in the frame of the shock must be small to ensure a low optical
depth for gammas.  Hence, $p-\gamma$ collisions are favored over 
$p-p$, although as we will show, even this preferred reaction places 
unreasonable energetic requirements on the burst environment.

Following \citet{Waxman97}, the energy loss rate due to pion
production is
\begin{equation}
\frac{1}{t_\pi}
   = \frac{1}{2 \Gamma_p^2} \int_{E_o}^\infty dE \sigma_\pi(E) \xi(E) E
\int_{E/2\Gamma p}^\infty \frac{d\epsilon}{\epsilon^{2}}\frac{d \phi_\gamma(\epsilon)}{d\epsilon} ~,
\end{equation}
where $E_0 \approx 0.15$ GeV  is the threshold for pion production.
   In the first integral, $\sigma_\pi$
is the cross section for
pion production due to a collision with a photon of energy $\epsilon_\gamma$
in the rest frame of the proton, and  $\xi(E)$ is the average fractional
energy lost
to the pion. The second integral is over the low energy GRB
spectrum, where $\phi_\gamma(\epsilon_\gamma)$  is the
photon flux in the frame of the proton.

The evaluation of $t_\pi$
can be simplified \citep{Waxman97} by integrating the
pion production cross section  over the
broken-power-law GRB spectrum (Eq. [\ref{batsespec}])
transformed back to the frame of the expanding plasma.
Approximating the integral over the pion production cross
section by the contribution from the peak of the
$\Delta$-resonance  \citep[as in][]{Waxman97} we deduce $t_\pi$ for a general
spectral power law index $\beta_i$:
\begin{equation}
\frac{1}{t_\pi} = \frac{c U_\gamma}{\epsilon_{\gamma
b}} \frac{\sigma_{peak} \xi_{peak}}{(1+\beta_i)} \frac{\Delta
E_{peak}}{E_{peak}} \times \mathrm{min} \left[1, \left( \frac{2 \Gamma_p
\epsilon_{\gamma b}}{E_{peak}} \right)^{\beta_i -1} \right]~~\mathrm{s}^{-1}~,
\label{tpi}
\end{equation}
where $\sigma_{peak}\approx 5 \times 10^{-28}$ cm$^2$ is the
$\Delta$-resonance cross section, while $E_{peak}=0.3$ GeV
and $\Delta E_{peak} \simeq 0.2$ GeV are the energy and
width of the resonance, respectively.  The fractional energy lost at
the peak is $\xi_{peak} \approx
0.2$, and  $U_\gamma$ is the photon energy density of {\it BATSE} gamma
rays in the frame of the
fireball, as before.

As before we estimate the fractional power radiated as 
\begin{eqnarray}
f_\pi & = & \frac{\Gamma \Delta t}{(1+z) t_\pi} \nonumber \\
  & = & \frac{4.5 \times
10^{-4}}{(1+\beta_i)} \frac{L^{ob}_{\gamma,51}
(1+z)^2}{\epsilon^{ob,{\rm MeV}}_{\gamma b} \Gamma^4_{300} \Delta t}
\begin{cases}
\times \left(\frac{E^{ob}_p}{E^{ob}_{pb,\pi}} \right)^{\beta_h -1}~, & \text{if
$E^{ob}_p < E^{ob}_{pb,\pi}$,} \\
\times 1~, & \text{if $E^{ob}_p \ge E^{ob}_{pb,\pi}$,}
\end{cases}
\label{fpiob}
\end{eqnarray}
where $\epsilon^{ob}_{\gamma b} \approx 1$ MeV is the break energy of
the two power laws of the observed GRB spectrum.  The last factor
in equation (\ref{fpiob})
   describes a break in the proton spectrum.
In the observer frame this break energy is \citep{Waxman97}
\begin{equation}
E^{ob}_{pb,\pi} = 1.3 \times 10^{7} \frac{
\Gamma^2_{300}}{(1+z)^2\epsilon^{ob,{\rm MeV}}_{\gamma b}}~~\mathrm{GeV} ~.
\end{equation}

Roughly half of the energy lost by the protons goes into
$\pi^+$'s, which quickly decay into neutrinos and positrons (through
$\mu^+$s).  In this work, we have ignored the effects of these decay
products on the emerging gamma-ray spectrum.  For neutrinos
this is reasonable since there is very little chance of them interacting
further.  The positrons, however, may influence the gamma-ray spectrum
through positron-synchrotron radiation \citep{bottcher} or pair
annihilation.

The other half of the energy lost by the protons goes into $\pi^0$'s, which
then decay
into two photons. The mean pion energy is $\xi_{peak} E_p$. When
the $\pi^0$ decays, the energy is shared equally among the
photons. Hence, each gamma ray has an average energy
\begin{equation}
E_\gamma = \xi_{peak} E_p/2~~.
\label{egammap}
\end{equation}
Now from equation (\ref{phigam}) 
\begin{eqnarray}
\left(\frac{d\Phi_\gamma}{dE_\gamma}\right)_\pi& = & f_\pi \left(\frac{2}{\xi_{peak}}
\right)^{2-\alpha} AE^{-\alpha}_\gamma \nonumber
\\
   & = & A^\pi_p D_\pi \times
\begin{cases}
(1+\beta_h)^{-1} \left( E^{ob,{\rm GeV}}_{\gamma b,\pi}
\right)^{1-\beta_h} \left( E^{ob,{\rm GeV}}_\gamma
\right)^{\beta_h -\alpha -1}~, & \text{if $E^{ob}_\gamma < E^{ob}_{\gamma
b,\pi}$,} \\
(1+\beta_l)^{-1} \left( E^{ob,{\rm GeV}}_\gamma
\right)^{-\alpha}~, & \text{if $E^{ob}_\gamma \ge
E^{ob}_{\gamma b,\pi}$,}
\end{cases}
\label{pispec}
\end{eqnarray}
where $A^\pi_p$ is a normalization constant and
\begin{equation}
D_\pi = 4.5 \times 10^{-4} \frac{L^{ob}_{\gamma,51}
(1+z)^{2-\alpha}}{\epsilon^{ob,{\rm MeV}}_{\gamma
b} \Gamma^4_{300} \Delta t} \left(
\frac{2}{\xi_{peak}} \right)^{2-\alpha}~.
\end{equation}
The observed break energy
$E_{\gamma b,\pi}^{ob}$ in the pion decay
gamma spectrum is given by
\begin{equation}
E^{ob}_{\gamma b,\pi} = \frac{\xi_{peak}}{2}E^{ob}_{pb,\pi} \approx 1.3
\times 10^{6} \frac{\Gamma^2_{300}}{(1+z)^2 \epsilon^{ob,{\rm MeV}}_{\gamma
b}}~\mathrm{GeV}~.
\end{equation}
Above this energy
the spectrum should obey the
$d\Phi_\gamma/dE_\gamma  \sim E_{\gamma}^{-2}$
of the protons and below this break
energy, the exponent should
be harder by one power,
i.e.~$d \Phi_\gamma/dE_\gamma \sim E_{\gamma}^{-1}$.
As a practical matter, photons with energy as high
as the break energy will not be observed, as they will be
extinguished by pair production as described below.  

\section{Photon, Proton, and Electron Luminosities at the Source}

\begin{figure}
\includegraphics[width=6in]{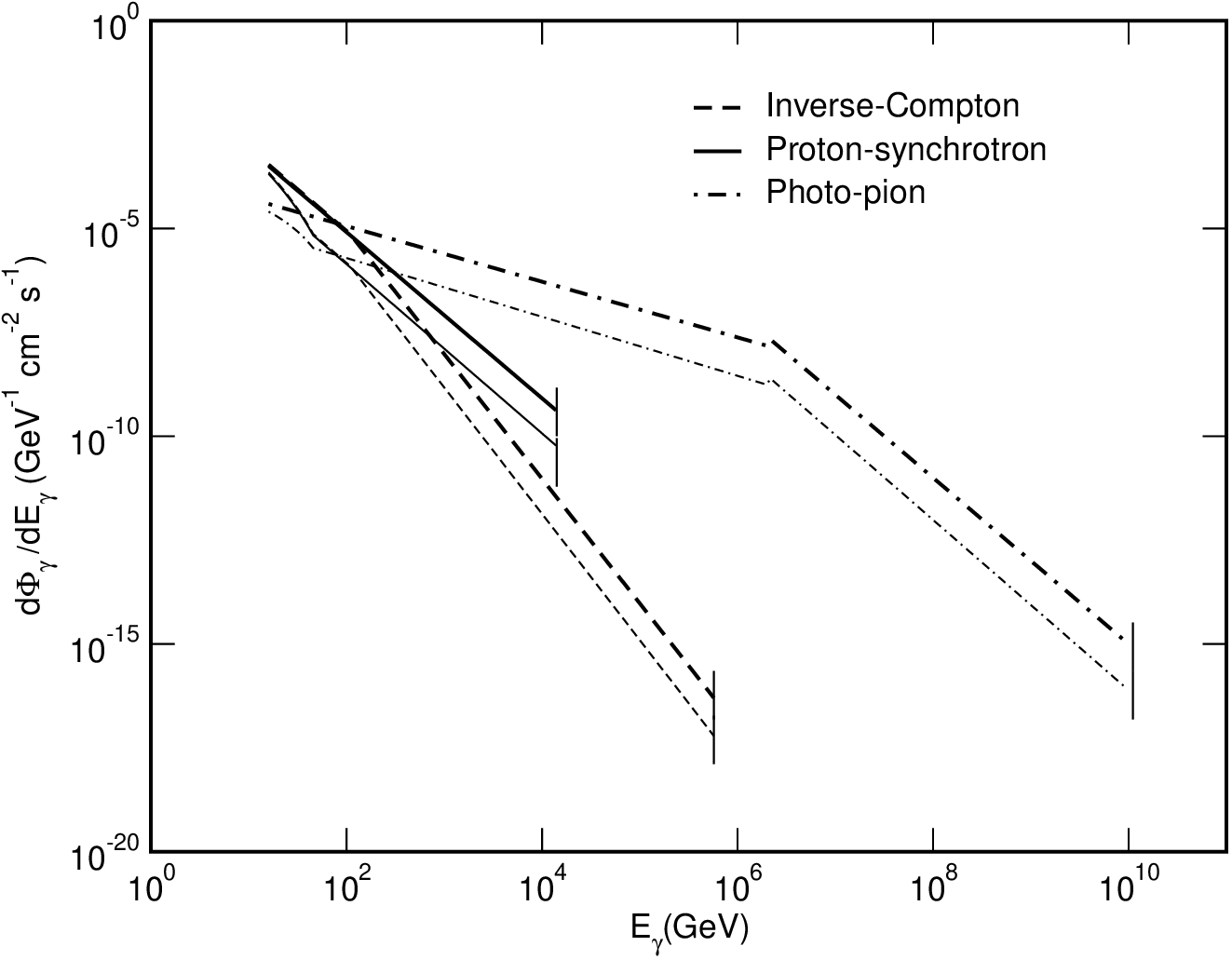}
\caption{Illustrative gamma-ray spectra for the three models discussed in the text.
Spectra have been normalized to produce the muon excess observed by Project {\it GRAND} for GRB 971110.  The curves begin at the detection threshold for
Project {\it GRAND} and run to the respective cut-offs of each model.
The thick lines are the raw source spectra.  
The thin lines illustrate the effects of internal pair-production optical 
depth on the source spectra.  The change in energy dependence of the internal
optical depth is apparent above about 50 GeV.
\label{figsource}
}
\end{figure}

\begin{figure}
\includegraphics[width=6in]{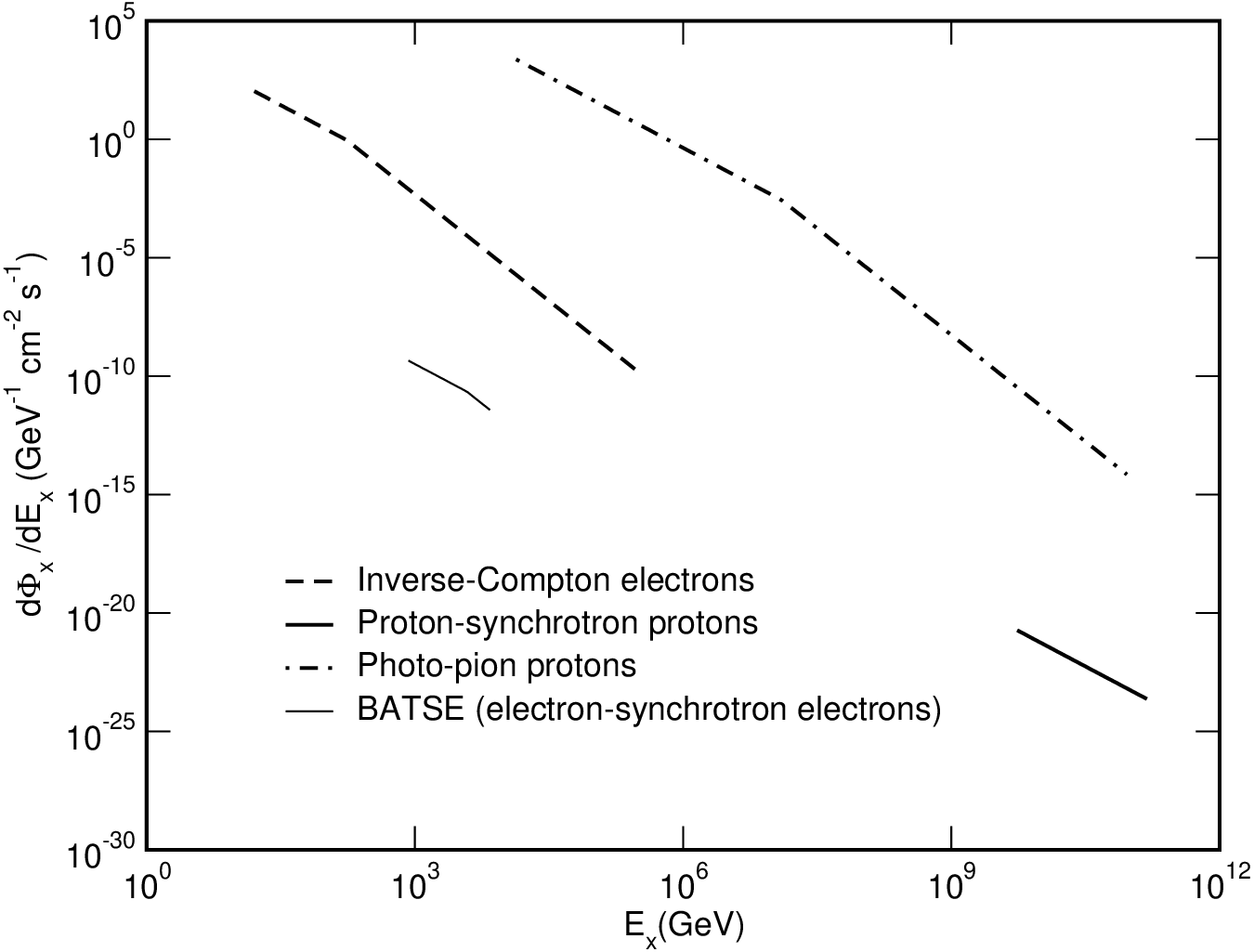}
\caption{Illustrative electron and proton spectra in the source frame 
for the three models discussed in the text.
Spectra have been normalized to produce the muon excess observed by Project {\it GRAND} or to fit the observed {\it BATSE} spectrum for GRB 971110.
\label{figpesource}
}
\end{figure}

  From the above it is clear that the three models considered here imply
different spectral shapes for the high-energy gamma component.
Figure \ref{figsource} compares the initial spectra 
 for all three mechanisms normalized to  reproduce the Project {\it GRAND} observations 
for GRB 971110 (as explained in Section 7).  This would correspond to the unrealistic limit
of no self or intergalactic absorption.  Nevertheless, this illustrates
the fact that
these mechanisms have significantly different energetic requirements.
Clearly, the most favorable energetically are 
the inverse-Compton and proton-synchrotron models.
Figure \ref{figpesource} further illustrates this point by reproducing the 
source proton and electron spectra required by the various models, again 
normalized to reproduce the Project {\it GRAND} observations.  
Also included in 
this figure is the source electron spectrum required to only produce the 
observed {\it BATSE} data for this burst, following the electron-synchrotron 
model.  The Project {\it GRAND} result, if it represents a real 
detection, requires a much higher flux of electrons 
to produce sufficient high-energy gamma rays through inverse-Compton 
scattering.  This is probably a troubling requirement for the inverse-Compton 
model.

To calculate these spectra we have used the normalization
of the gamma-ray spectrum from the observed muon excess to determine
the normalization of
the associated ultra-relativistic proton and electron spectra.
This is straightforward for the proton-synchrotron and photo-pion models, 
since the proton normalization appears explicitly in our final 
expressions (equations [\ref{pispec}] and [\ref{synspec}]).  For the 
inverse-Compton model, we have followed \citet{SE01} to estimate 
the electron normalization from the following approximate relation 
between the low-energy electron-synchrotron spectrum, assumed to be 
observed by {\it BATSE}, and the inverse-Compton spectrum in the energy 
range of Project {\it GRAND}
\begin{equation}
E_\gamma \left( \frac{d\Phi_\gamma}{dE_\gamma} \right)_{\rm IC} \simeq
0.5 r_d \sigma_T n \epsilon_\gamma \left( \frac{d\phi_\gamma}
{d\epsilon_\gamma} \right)_{\rm e-s} ~.
\label{eq:ICes}
\end{equation}
We take the electron number density $n$ to be 1 cm$^{-3}$.  The 
electron-synchrotron spectrum is assumed to have a form 
directly comparable to equation (\ref{synspec}).
We evaluate 
equation (\ref{eq:ICes}) at $E_\gamma=E^{ob}_{\rm cool}$ and 
$\epsilon_\gamma = \epsilon^{ob}_{\gamma b}$ to find $A_e^{\rm e-sync}$.

On the other hand, if one attributes \citep{Waxman95} the observed cosmic-ray
excess
above $10^{20}$ eV to the energetic protons accelerated in GRBs,
then an independent estimate of the proton normalization at the source 
can be obtained.
Following \citet{Waxman95} we note that the observed cosmic ray
flux above $10^{20}$ eV is $\sim 3 \times 10^{-21}$  cm$^{-2}$
s$^{-1}$ sr$^{-1}$ \citep{Bird}, corresponding to an average universal
   number density of energetic cosmic rays of
$n_{CR} \sim 10^{-30}$ cm$^{-3}$.  If this density is due
to GRBs then the number of protons with energies greater than 10$^{20}$
eV produced per GRB must be
\begin{equation}
N(E_p > 10^{20}) = n_{CR} /\nu_\gamma \tau_{CR} \sim 5 \times 10^{44} ~~,
\end{equation}
where $\nu_\gamma \approx 2 \times 10^{-10} (h/70)^3$ Mpc$^{-3}$ yr$^{-1}$
is the cosmological GRB rate \citep{tot97a, tot99, Schmidt}, 
and $\tau_{CR} \approx 3
\times 10^8$ yr is the lifetime of protons with $E_p > 10^{20}$ eV.
With our nominal $E_p^{-2}$ spectral form, we deduce an
absolute normalization of the
proton spectrum emerging from an average GRB source of
$dN_p/dE_p = 5 \times 10^{55} E_{\rm GeV}^{-2}$ GeV$^{-1}$.
This is to be compared with the number of photons
with $E > 1$ MeV emerging from a
   nominal GRB.  If we assume that $10^{53}$ erg is released in gammas
above 1 MeV for an energetic photon
spectrum of $d\Phi_\gamma/d E_\gamma \propto E_\gamma^{-2}$, then the
normalized spectrum for gammas above 1 MeV would be
$dN_\gamma/dE_\gamma = 6 \times 10^{58} E_{\rm GeV}^{-2}$ GeV$^{-1}$.
Thus, one expects about 1200 gammas per proton from such a burst.

Since both the proton-synchrotron and photo-pion models
are based upon the same underlying baryon content in the relativistic
plasma, it is instructive to summarize the relative efficiency 
of these two 
mechanisms for generating energetic gamma rays in GRBs.  Let us consider
a typical burst with $\beta_h=2$ and a proton spectrum with
$\alpha=2$.  Then in the region $E^{ob}_\gamma \sim 1$ TeV, the
fractional energy loss into these two mechanisms is
\begin{equation}
f_\pi = 1.7 \times 10^{-7}
\frac{L^{ob}_{\gamma,51}(1+z)^4}{\Gamma^6_{300} \Delta t} \left(
\frac{E^{ob}_\gamma}{\mathrm{TeV}} \right) ~,
\end{equation}
and
\begin{equation}
f_\mathrm{p-sync} = 
\mathrm{min} \left\{1, 
\left[6.7 \times 10^2 
\frac{\xi_B^{3/4}(L^{ob}_{\gamma,51})^{3/4} (1+z)^{5/2}}{\Gamma^4_{300} \Delta t^{1/2}}
\left(\frac{E^{ob}_\gamma}{\mathrm{TeV}} \right)^{1/2} \right] \right\}~.
\end{equation}
The factor of $1.7 \times 10^{-7}$ suggests that
proton-synchrotron emission will dominate over the photo-pion
production for the parameters and gamma-ray energy considered.  
Only for sufficiently small $\xi_B$ ($\lesssim 10^{-4}$) does 
the efficiency of the proton-synchrotron mechanism begin to deviate 
from unity as $\xi_B^{3/4}$.

A direct comparison between inverse-Compton scattering 
and proton-synchrotron emission is not possible since these two mechanisms
rely upon different underlying energy sources, i.e. relativistic electrons 
for inverse-Compton
scattering and relativistic protons for 
proton-synchrotron emission.  Nevertheless, we note that both of these methods can be very
efficient.  In the region $E^{ob}_\gamma \sim 1$ TeV, 
$f_{\rm IC} \sim 1$ and $f_\mathrm{p-sync} \sim 1$ for the range of parameters 
considered.  This suggests that inverse-Compton 
scattering provides a very competitive
mechanism for the generation of energetic gammas.  Furthermore, it makes no 
requirement on the pre-existence of a magnetic field or baryon loading
in the  fireball plasma.

Another important difference among all of these mechanisms is their
associated cut-off energies.  As noted above, there should be a
cut-off for the proton-synchrotron
spectrum around $E^{ob,max}_{\gamma,p-s} \approx 15$ TeV,
 corresponding to
a cut-off in the ultra-relativistic proton energies.  The inverse-Compton 
spectrum has a much larger cut-off at around $E^{ob,max}_{\gamma,IC} \approx 
400$ TeV, corresponding to a cut-off in the relativistic electron energies.
For pion decay, however,
 the spectrum may extend all the way to $10^7$ TeV \citep{Waxman97}, but
with a break at around 1300 TeV.  These differences have  a large effect 
on the implied total source luminosities of the
bursts, as is apparent in Figure \ref{figsource}.

\section{Pair Production Optical Depth}

The spectra derived above must be corrected for two effects, both of which
are due to pair production by energetic photons.  First, within the burst
environment, energetic
gamma rays will interact with other photons to produce $e^+ - e^-$
pairs.  If this process is
highly efficient, $\sim$ TeV gamma rays may not be able to
escape from the burst.  Even if some photons escape, this
self-absorption will affect the implied source luminosity.

The second effect is due to absorption along the line-of-sight
from the burst environment.  Here the energetic gamma rays
interact with the intergalactic infrared and microwave
backgrounds. This effect can cause a dramatic shift in the
spectrum of energetic gammas in the TeV range
depending upon the distance to the burst.

\subsection{Internal Optical Depth from Pair Production}
A photon of energy $E_\gamma$ interacts mainly with target photons of
energy $\epsilon_\gamma \sim 2 m_e^2 c^4/E_\gamma$ in the shock frame.  We
can approximate the cross section for pair-production as $3 \sigma_T /16$,
where $\sigma_T = 6.6 \times 10^{-25}$  cm$^2$ is the Thomson cross
section.  Then the optical depth can be approximated as
\begin{equation}
\tau_{\gamma \gamma,int} \sim \frac{3}{16} \sigma_T  \frac{D_z^2}{r_d^2} 
\epsilon_\gamma \frac{d
\phi_\gamma [\epsilon_\gamma /(1+z)]}{d \epsilon_\gamma} \frac{\Delta t}
{\Gamma(1+z)} ~,
\label{tauggeq}
\end{equation}
where $\Delta t \Gamma/(1+z)$ is the width of the emitting region as measured in
the shock frame.

This formula is similar to that of \citet{Waxman97}, 
except that this form takes into
account the spectral break in the low-energy GRB photons.  This break is
important since it implies that the optical depth is proportional to $\sim E_\gamma$ for
$E^{ob}_\gamma \lesssim 2 m_e^2 c^4 \Gamma^2 /\epsilon^{ob}_{\gamma
b}\approx 50$ GeV, but roughly constant for higher energies.  
The \citet{Waxman97}
result is only valid in the lower energy range.  The proper energy
dependence is important because the internal optical depth can be
of order unity.  In the case of GRB 971110, the internal optical depth
at 100 GeV is $\approx 3$ 
implying that some energetic gammas could emerge.  
The thin lines in Figure \ref{figsource} show how the three 
source spectra normalized to fit GRB 971110 would be modified
by internal absorption.
The change in the energy dependence of the internal optical depth 
is apparent above about 50 GeV.
For the remainder of the 
bursts, we were unable to place meaningful constraints 
on the internal optical depth due to large uncertainties in 
the fit parameters, particularly 
$\beta_h$ and $\Gamma$.  For these bursts, we have taken the very 
optimistic assumption of $\tau_{\gamma\gamma,int} =0$.

\subsection{Intergalactic Optical Depth from Pair Production}

Another important constraint on the observed burst spectra
comes from the absorption of photons via pair production
during collisions with the intergalactic infrared background.  In this work
we use a calculated optical depth for
intergalactic absorption based upon the standard formulation
\citep[e.g.][]{Salamon}.  We use a model for
the luminosity evolution of background light from
\citet{tot97b}.  The dust emission  component
is calculated assuming a dust emission spectrum similar
to that of the Solar neighborhood.  The fraction of light
absorbed by dust is adjusted to reproduce the observed
far infrared background from COBE \citep{Hauser}.
This method is summarized in \citet{tot00}.  It is consistent with other optical-depth
calculations \citep[cf.][]{Salamon, Primack}.

The resultant $e^+ - e^-$ pairs may then further modify the original gamma-ray 
spectrum by providing a medium for some of the remaining gamma rays to undergo 
intergalactic inverse-Compton scattering.  This process results in complicated 
showers of secondary electrons and gammas.  These secondary gammas may be observed, 
but over a much longer time-scale, since much of this secondary light traverses 
a longer path length.  The flux from these secondary gamma rays is probably below 
the detection threshold of current arrays such as Project {\it GRAND}.  
Nevertheless, this might be a noteworthy effect.

\begin{figure}
\includegraphics[width=6in]{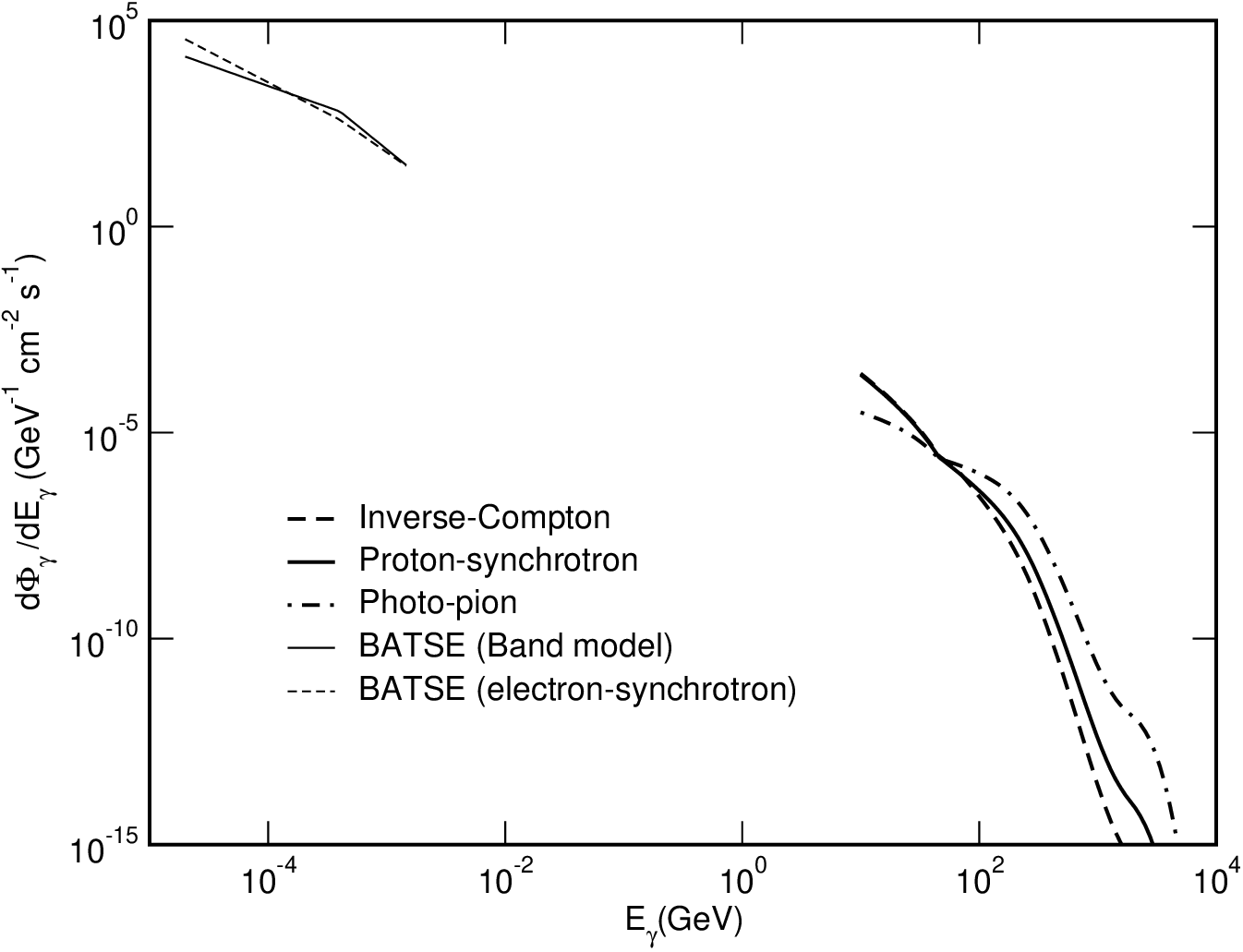}
\caption{Illustration of the effects of internal and intergalactic
 pair-production optical depth
on the source spectra shown in Figure \ref{figsource}.  This calculation assumes that the source (GRB 971110)
 is at a redshift of $z = 0.6$.  The change in energy dependence of the 
internal optical depth is apparent above about 50 GeV.  For illustration, 
this plot also includes the observed {\it BATSE} spectrum for this burst, fit 
with both the Band (Eq. [\ref{batsespec}]) and electron-synchrotron 
models.
\label{figobs}
}
\end{figure}

Figure \ref{figobs} 
shows final spectra in the observer's frame 
when both internal and intergalactic
absorption are taken into account, assuming the burst is 
arriving from a redshift of $z = 0.6$ 
(appropriate for GRB 971110).
Even though the source spectra are
vastly different, the observed high-energy gamma spectra are quite similar.
Hence, the implied source energy requirement may be the only way to 
distinguish between the models.  For illustration, 
Figure \ref{figobs} also includes the observed 
{\it BATSE} spectrum for this burst, fit 
with both the Band (Eq. [\ref{batsespec}]) and electron-synchrotron 
models.

\section{Results}
\label{sec:results}

In this work we use the muon observations of Project {\it GRAND} 
to fix the
normalization  (or upper limit) in each of the models described above
for the various bursts analyzed. That is, the
spectral shape is fixed from equations (\ref{pispec}), (\ref{synspec}), or
(\ref{icspec}), and the number of muons expected is then
computed using
\begin{equation}
N_\mu = dA \times \mathrm{T}90 \times \int^\infty_{E_{min}} dE^{ob}_\gamma
P_\mu (E^{ob}_\gamma)
\frac{d\Phi_\gamma (E^{ob}_\gamma)}{dE^{ob}_\gamma}
 \exp(-\tau_{\gamma \gamma})~,
\label{nmueq}
\end{equation}
where $dA$ is the collecting area of the Project {\it GRAND} array 
(the effective
area at the time of GRB 971110 was approximately $6.3 \times 10^5$ 
cm$^2$), $E_{min}$ is the primary gamma-ray detection threshold for Project 
{\it GRAND} ($\sim 10$
GeV), and $P_\mu \approx 7.0 \times 10^{-5} (E_\gamma^\mathrm{GeV})^{1.17}$ is the
probability per primary for a muon to reach detection level at 
Project {\it GRAND}.  
This
probability (valid for $E_\gamma \gtrsim 3$ GeV) was computed by 
\citet{Fasso} using the
Monte Carlo atmospheric absorption code, {\it FLUKA}.  
Here $\tau_{\gamma \gamma}$ includes both the internal and
intergalactic optical depth estimated for each burst as described above. 
For illustration, Table \ref{optdeptab} summarizes 
some estimates of the relative magnitudes of
internal and intergalactic optical depths at two energy scales, 100 GeV
and 1 TeV.

In practice, the
integral in equation (\ref{nmueq}) is cut off at $\approx 30$ TeV since the optical depth is quite
high for photons above this energy.  The normalization constants for each
model are then adjusted so that $N_\mu$ agrees with the $2 \sigma$ upper
limits set by Project {\it GRAND}, except for GRB 971110 
where we used the observed mean value (see Table \ref{BATSETab}).  
Here we adopt typical values for the
degree of equipartition $\xi_B$ and the relativistic Lorentz
factor $\Gamma$, namely $\xi_B \approx 1$ and $\Gamma \approx 300$.  Below we explore the dependence of our results on a broad range of these parameters.

We can then use our normalized gamma spectra to estimate the total energy
emitted in
high-energy gammas at the source.  For the inverse-Compton model, 
we can also estimate the energy emitted in 
electrons, noting again the characteristic spectrum of a Fermi mechanism
given in equation (\ref{fermi}).  Similarly we can estimate the energy 
emitted in protons for the proton-synchrotron and photo-pion models.  
Implied energies for photon 
emission into $4 \pi$ are given in Tables \ref{ictab}, \ref{psynctab}, and 
\ref{photoptab} for the
inverse-Compton, proton-synchrotron, and photo-pion models,
respectively.  We also list the required energies of the source protons 
or electrons, as appropriate.  Our results assume $\alpha=2$ for the accelerated 
electron and proton spectra.  For GRB 971110 we estimated the statistical uncertainties of our results using Monte Carlo techniques to explore the parameter space of each of the models assuming Gaussian error distributions.

As evidenced by the large uncertainties, the models are not well constrained at
 present.  Nevertheless, several points are worth noting from the tables.  
For the most significant possible detection (GRB 971110), 
the energetic requirements for the IC model 
($E_\gamma^{Tot,{\rm IC}} = (2.6 \pm 8.6) \times 10^{55}$
erg) and the proton-synchrotron model 
($E_\gamma^{Tot,p-sync} = (3. \pm 10.) \times 10^{55}$ erg) are much less than that
for a photo-pion mechanism ($E_\gamma^{Tot,\pi} = (4. \pm 27.) \times 10^{62}$
erg), as expected.

The distinction between the inverse-Compton and proton-synchrotron 
mechanisms has an important consequence on the magnetic field of the 
GRB, specifically the degree of equipartition, $\xi_B$.  In the case 
of inverse-Compton scattering, the ratio of the inverse-Compton 
luminosity $L_\gamma^{\rm IC}$ to electron-synchrotron luminosity 
$L_{\rm sync}$ 
should be equal to the ratio of the IC target photon energy density 
$U_\gamma$ to the magnetic field energy density $U_B$ since both
mechanisms arise from the same population of hot relativistic electrons.  Thus,
\begin{equation}
\frac{L_\gamma^{\rm IC}}{L_{\rm sync}} = \frac{U_\gamma}{U_B} ~,
\label{lratio}
\end{equation}
where $U_B=\xi_B U$ and $U$ is the total rest-frame energy density of 
the emission region.  If we adopt the generally accepted view that the $\sim$MeV
gamma rays seen by {\it BATSE} are caused by electron-synchrotron radiation
then $L_{\rm sync} = L_\gamma$.  For GRB 971110, $L_\gamma^{\rm IC}/L_\gamma \sim 10^4$, which implies
\begin{equation}
\xi_B U \lesssim 10^{-4} U_\gamma~.
\label{xibu}
\end{equation}

Since the energy density of the $\sim$MeV {\it BATSE} gamma rays is likely to be
higher than any other radiation source available for inverse-Compton
scattering, we identify this energy density as $U_\gamma$.  This association
 has been implicit throughout this paper.  On the other hand, since 
$L_\gamma^{\rm IC} \sim 10^4 L_\gamma$ the energy density of the $\sim$TeV
gamma rays must be greater than $U_\gamma$ by about $10^4$.  We can then
use $10^4 U_\gamma$ as a lower limit for the total rest-frame energy
density $U$.  Combining this with Eq. (\ref{xibu}) we get the following
 upper limit on the degree of equipartition
\begin{equation}
\xi_B \lesssim 10^{-8}~.
\end{equation}
Note also that Eq. (\ref{lratio}) holds only in the classical regime.  
In the KN regime $\xi_B$ is even smaller 
because inverse-Compton scattering becomes less efficient.
Hence, the inverse-Compton model is at odds with models that propose
GRBs as the source of ultra-high energy cosmic rays, since those models
require a magnetic field near equipartition, $\xi_B \sim 1$ 
\citep[cf.][]{Waxman95}.

\begin{figure}
\includegraphics[width=6in]{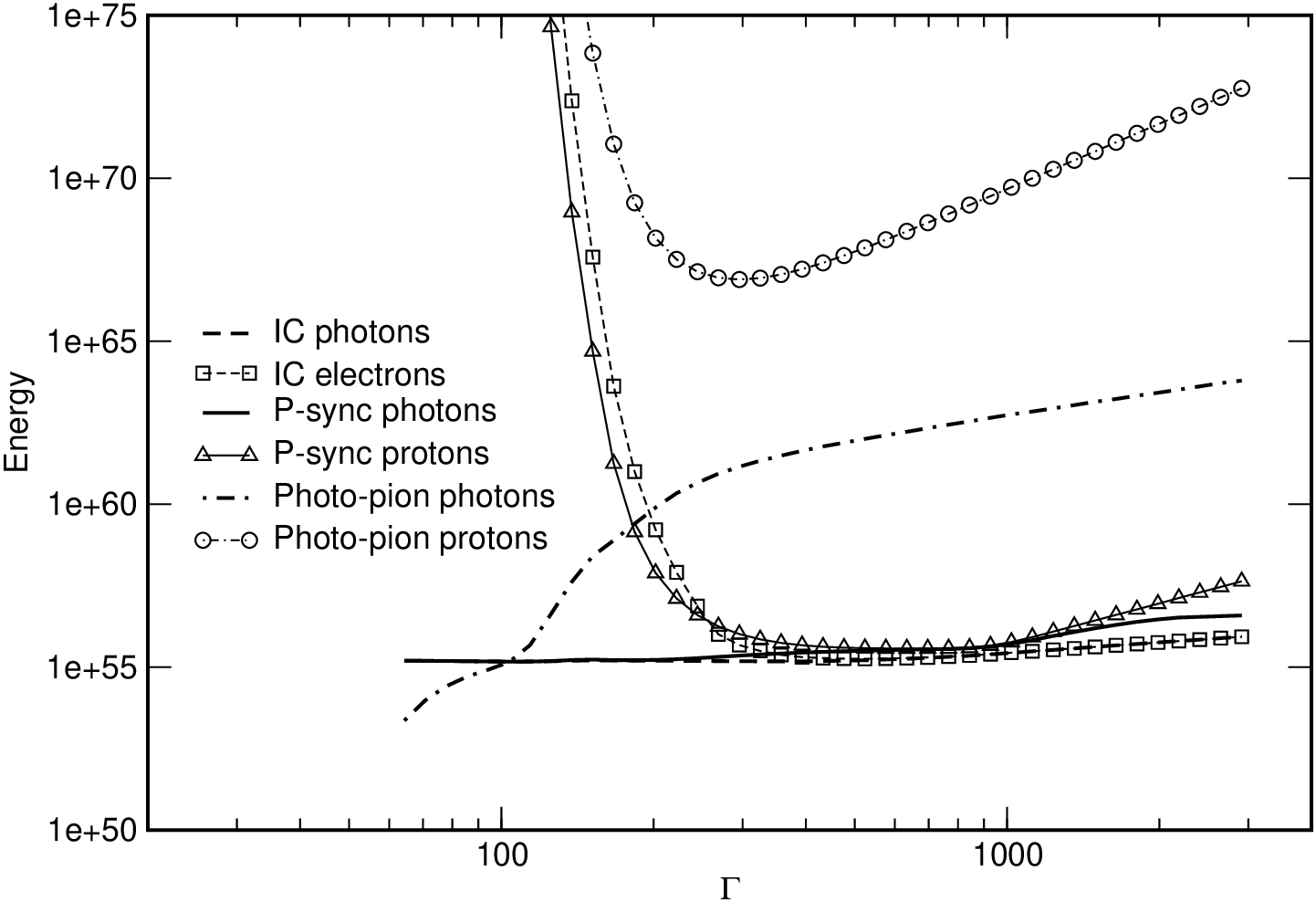}
\caption{Plot of the energy escaping in gamma rays (accounting for $\tau_{\gamma \gamma,int}$) and the source energy in protons or electrons for the inverse-Compton, proton-synchrotron, and photo-pion models as a function of the Lorentz boost of the GRB fireball.  These curves correspond to spectra that were normalized to fit the observed muon excess for GRB 971110.  The sharp increase in the energy content of the protons and electrons 
at small $\Gamma$ enters primarily through the internal optical depth.  
\label{figGam}
}
\end{figure}
\begin{figure}
\includegraphics[width=6in]{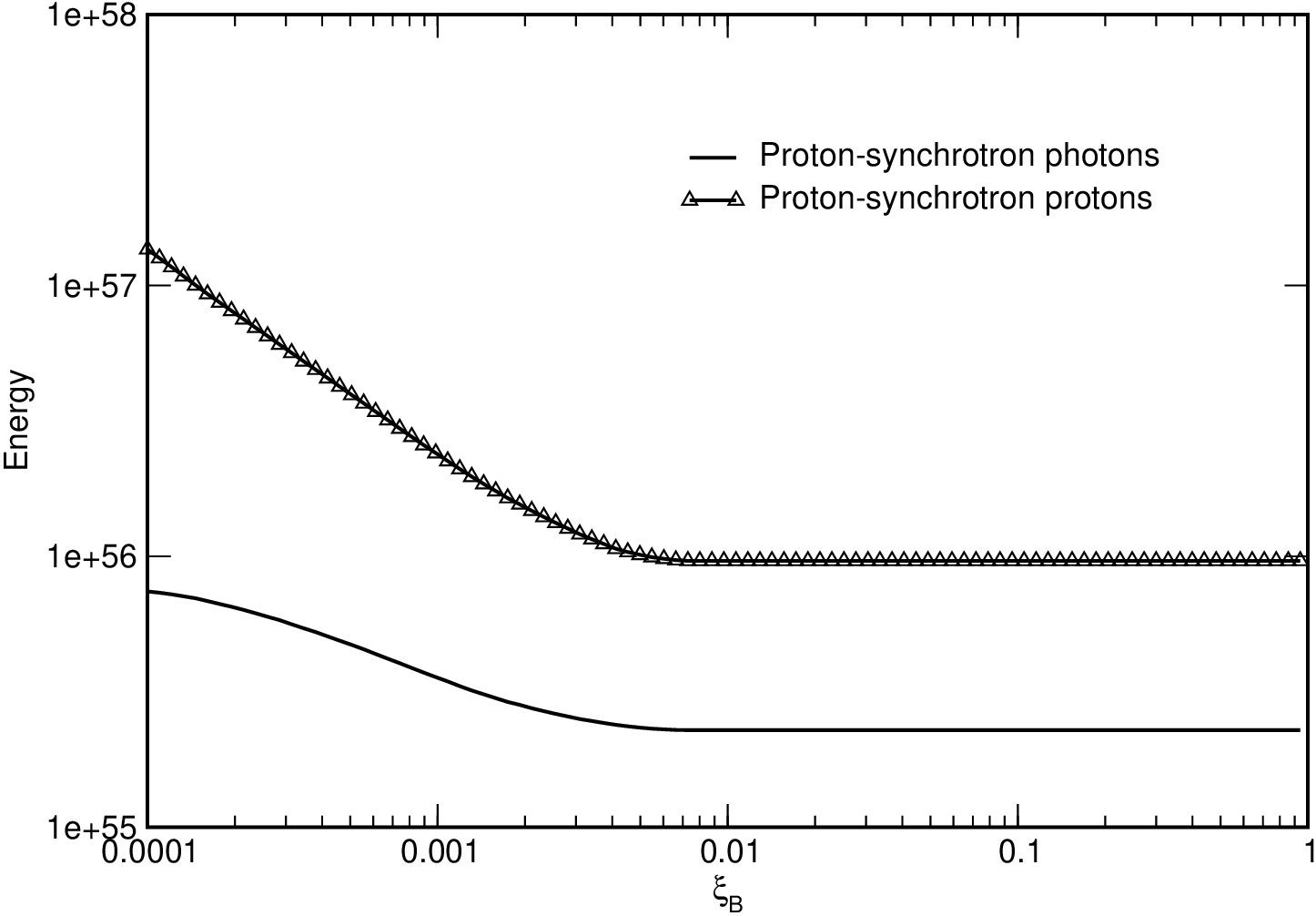}
\caption{Plot of the energy escaping in gamma rays (accounting for $\tau_{\gamma \gamma,int}$) and the source energy in protons for the proton-synchrotron model as a function of the magnetic equipartition of the GRB fireball.  These curves correspond to spectra that were normalized to fit the observed muon excess for GRB 971110.  The offset between the photon and proton spectra for $\xi_B \gtrsim 0.01$ is due solely to absorption due to the internal optical depth, $\tau_{\gamma \gamma,int}$.
\label{figxi}
}
\end{figure}
\begin{figure}
\includegraphics[width=6in]{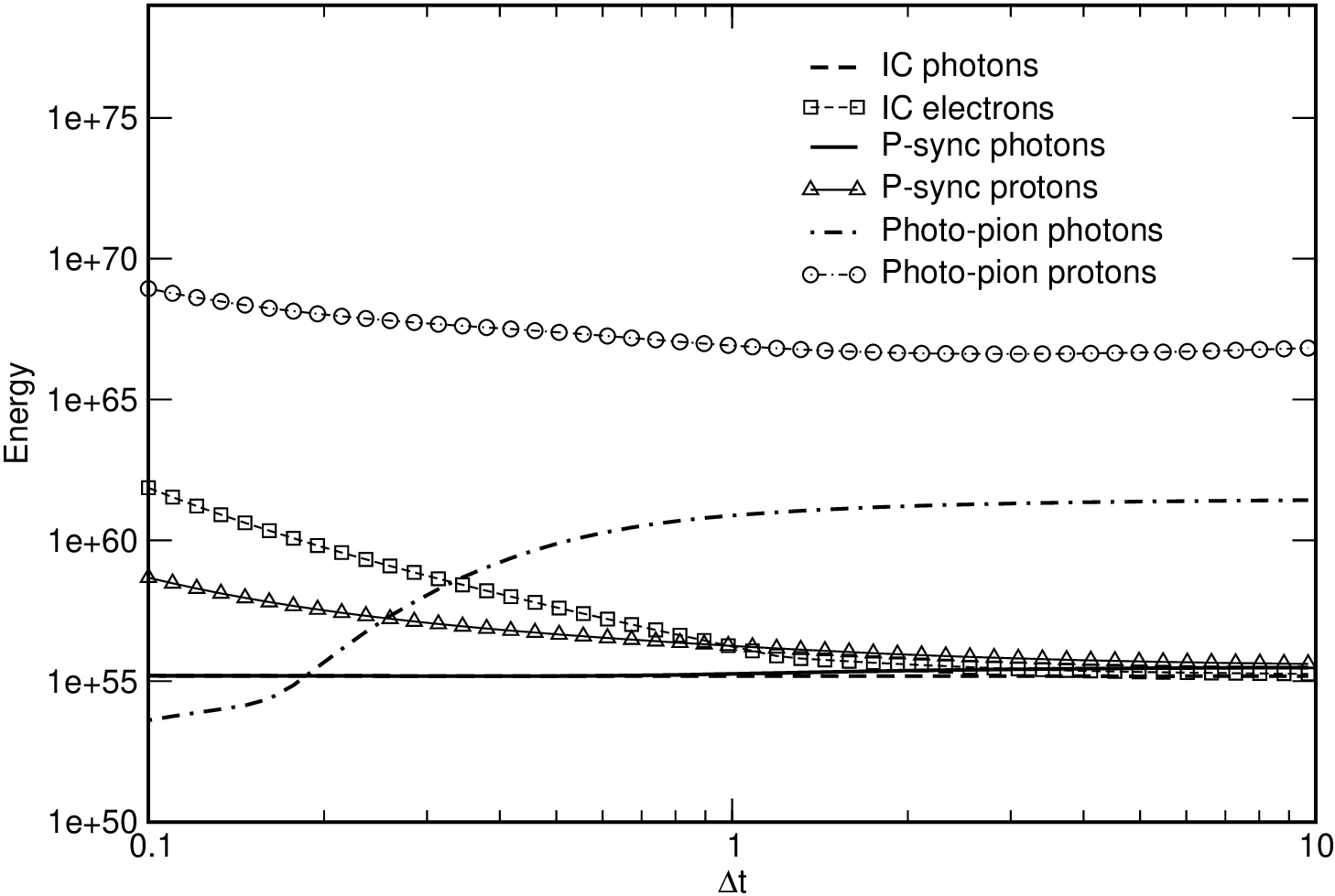}
\caption{Plot of the energy escaping in gamma rays (accounting for $\tau_{\gamma \gamma,int}$) and the source energy in protons or electrons for the inverse-Compton, proton-synchrotron, and photo-pion models as a function of the time variability of the GRB fireball.  These curves correspond to spectra that were normalized to fit the observed muon excess for GRB 971110.
\label{figdt}
}
\end{figure}

In Figures \ref{figGam}, \ref{figxi}, and \ref{figdt} 
we explore the dependence of our results upon 
a broad range of possible values for $\Gamma$, $\xi_B$, and 
$\Delta t$, respectively.  
These uncertainties were not formally included in our statistical estimates.  
Nevertheless, these figures support our qualitative conclusions.  
Specifically, the energetic requirements of all three models are in 
excess of that for the GRB for a broad range of parameters.  
Indeed most parameter variations in Figures \ref{figGam} - \ref{figdt} 
only exacerbate this problem.  However, 
the inverse-Compton and proton-synchrotron mechanisms 
are generally less sensitive to 
changes in $\Gamma$ over the range $100 \lesssim \Gamma \lesssim 1000$ 
or changes in $\Delta t$ over the range $0.1 \lesssim \Delta t \lesssim 10$.  
The sharp increase in the energy content of the protons and electrons 
at small $\Gamma$ enters primarily through the internal optical depth.  
Figure \ref{figxi} shows that the proton-synchrotron mechanism is highly
efficient over a fairly broad range of $\xi_B$ ($\gtrsim 0.01$).

\begin{figure}
\includegraphics[width=6in]{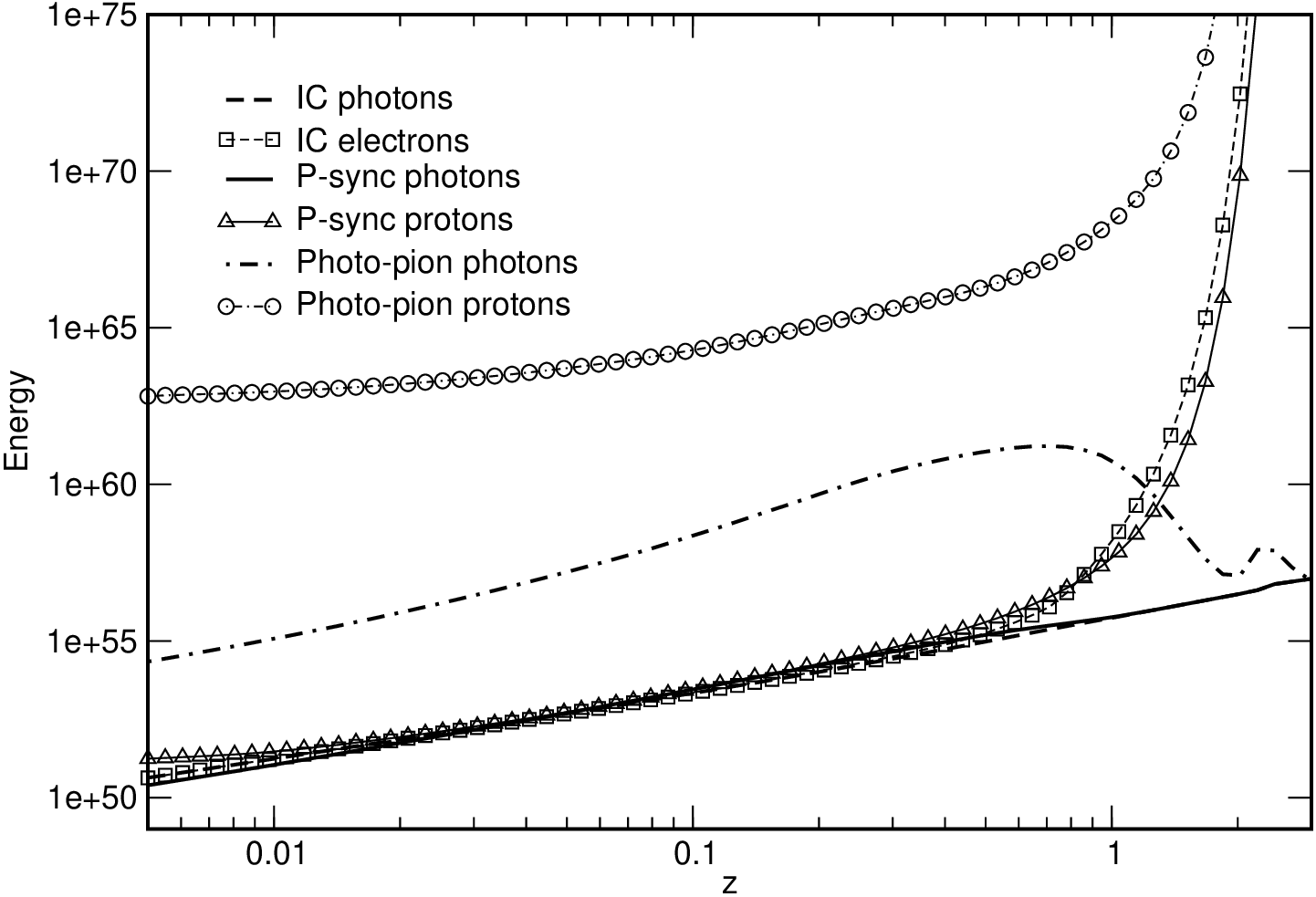}
\caption{Plot of the energy escaping in gamma rays (accounting for $\tau_{\gamma \gamma,int}$) and the source energy in protons or electrons for the inverse-Compton, proton-synchrotron, and photo-pion models as a function of the redshift of the GRB fireball.  These curves correspond to spectra that were normalized to fit the observed muon excess for GRB 971110.
\label{figz}
}
\end{figure}

Regardless of the mechanism, if GRB 971110 is indeed a detection and 
our estimated redshift of $z=0.6$ is valid, 
the implied energy in energetic ($E_\gamma > 1$ GeV) gamma rays 
is more than 100 times
higher than the energy in the low-energy gamma rays.  
This renders the already challenging energetic
requirements on the GRB source engine to be even more difficult.  
It may be possible 
to alleviate this difficulty if, perhaps, the photon beaming angle 
is much narrower for the high-energy component than for the 
low-energy GRB.  Another possibility is that the redshift is much 
lower than our estimated value for this burst.  
In Figure \ref{figz} we explore the energetic requirements 
for GRB 971110 over a very broad 
range of redshifts from 0.005 to 3.  This range is 
consistent with most currently 
measured GRB redshifts.  This figure illustrates quite clearly 
the critical role 
that intergalactic absorption plays in driving up the energetic 
requirements and highlights the need for accurately determining 
the true redshifts of GRBs.

\section{ Conclusion}

We have analyzed the eight GRBs discussed in \citet{Poirier}, 
which occurred above the Project {\it GRAND} array.
We have studied the implied energies in energetic ($\sim$TeV)
gamma rays (and associated electrons and protons) of such GRBs
in the context of three possible mechanisms:  
inverse-Compton scattering, proton-synchrotron radiation, and
photo-pion production.  Our analysis suggest that all of these 
models face significant energetic requirements.  Gamma-ray production by 
either inverse-Compton scattering or 
proton-synchrotron radiation is probably the most 
efficient process.  

Although it can not be claimed that TeV gammas have unambiguously 
been detected in
association with low-energy GRBs, we have argued that there is enough 
mounting evidence
to warrant further study.  Furthermore, we have shown that if 
TeV gammas continue to be observed, then they present 
some interesting dilemmas for GRB physics.  
In view of their potential as a probe of the GRB source environment, 
we argue that further efforts to 
measure energetic gamma rays in association with low-energy gamma-ray 
bursts are warranted.

\begin{acknowledgments}
The authors would like to thank M. S. Briggs for providing fits to these {\it BATSE} data.
This research has made use of data obtained from the High Energy Astrophysics
Science Archive Research Center, provided by NASA's Goddard
Space Flight Center.
Project {\it GRAND}'s research is presently being funded through a grant from the
University of Notre Dame and private grants.
This work was supported in part by DoE Nuclear Theory grant
DE-FG02-95ER40934 at the University of Notre Dame.  
This work was performed in part under the auspices of the US Department of 
Energy by the University of California, Lawrence Livermore National Laboratory under 
contract W-7405-Eng-48. One of the authors (TT) wishes to acknowledge support
under a fellowship for research abroad provided by
the Japanese Society for the Promotion of Science.  
\end{acknowledgments}

%
%

\clearpage

%
%

\clearpage

\begin{table}
\footnotesize
\caption{Project {\it GRAND}'s Response to Selected {\it BATSE} Bursts
\label{BATSETab}}
\begin{ruledtabular}
\begin{tabular}{ccccccc}
GRB &
Trig &
T90\footnotemark[1] &
RA\footnotemark[1] &
Dec\footnotemark[1] &
$\delta\theta$\footnotemark[1] &
$N_{\mu}$\footnotemark[2] \\

\hline

971110 & 6472 & 195.2 & 242 & 50 & 0.6 & $467 \pm 171$ \\
990123 & 7343 & 62.5 & 229 & 42 & 0.4 & $< 75$ \\
940526 & 2994 & 48.6 & 132 & 34 & 1.7 & $< 76$  \\
980420 & 6694 & 39.9 & 293 & 27 & 0.6 & $< 133$ \\
960428 & 5450 & 172.2 & 304 & 35 & 1.0 & $< 213$ \\
980105 & 6560 & 36.8 & 37  & 52 & 1.4 & $< 107$  \\
980301 & 6619 & 36.0 & 148 & 35 & 1.3 & $< 150$ \\
970417a & 6188 & 7.9 & 290\footnotemark[3] & 54\footnotemark[3] &
0.5\footnotemark[3] & $20 \pm 17$       \\

\end{tabular}
\end{ruledtabular}
\footnotetext[1]{Angles RA, Dec, and $\delta\theta$ in degrees and T90
in seconds.}
\footnotetext[2]{Upper limits are $2 \sigma$ confidence level.}
\footnotetext[3]{RA, Dec, and $\delta \theta$ for GRB 970417a are based
upon the {\it Milagrito} data.}

\end{table}

\clearpage

\begin{table}
\scriptsize
\caption{Observed and Inferred GRB Properties in {\it BATSE} Energy
Range \label{gammaraytab}}
\begin{ruledtabular}
\begin{tabular}{cccccc}
  &
$a$ &
$\beta_l$ &
$\beta_h$ &
$\epsilon^{ob}_{\gamma b}$ &
$\Delta t$ \\
GRB &
(cm$^{-2}$ s$^{-1}$ MeV$^{-1}$) &
  &
  &
(MeV) &
(s) \\

\hline

971110 & $0.0751\pm0.0095$ & $1.02\pm0.04$ & $2.33\pm0.11$ &
$0.404\pm0.041$ & 1.8 \\
990123 & $1.93\pm0.14$ & $0.60\pm0.01$ & $3.11\pm0.07$ & $1.29\pm0.07$ &
4.1 \\
940526 & $\le 0.217$ & $1.01\pm0.02$ & $>3.2$ &
$>1.28$ & 1.2 \\
980420 & $0.0815\pm0.0105$ & $0.18\pm0.08$ & $2.57\pm0.27$ &
$0.553\pm0.033$ & 1.8 \\
960428 & $0.0610\pm0.0086$ & $0.58\pm0.09$ &
$2.49\pm0.24$ & $0.433\pm0.033$ & 0.7 \\
980105 & $0.0332\pm0.0081$ & $0.69\pm0.06$ &
$2.82\pm0.19$ & $0.286\pm0.031$ & 1.2 \\
980301 & $0.0068\pm0.0020$ & $0.54\pm0.20$ &
$3.26\pm1.41$ & $0.317\pm0.033$ & 2.3 \\
970417a & $0.0100\pm0.0061$\footnotemark[1] & $1.0\pm0.5$\footnotemark[2] & $2.25\pm0.75$\footnotemark[2] & $0.250\pm0.150$\footnotemark[2] & 1.1 \\

\end{tabular}
\end{ruledtabular}
\footnotetext[1]{Spectral fits were not available for GRB 970417a.  
Properties are based upon assumed redshift $z \approx 0.7$, observed {\it BATSE} fluence, and average GRB spectral shape.}
\footnotetext[2]{Average parameters for all bright GRBs considered in Preece et al. (2000).}

\end{table}

\clearpage

\begin{table}
\footnotesize
\caption{Variability-Luminosity Redshift Estimates \label{zTab}}
\begin{ruledtabular}
\begin{tabular}{cccccc}
  &
$V$ &
$P_{256}$\footnotemark[1] &
$z$\footnotemark[2] &
$L_{256}$\footnotemark[3] &
$L_\gamma$\footnotemark[3] \\
GRB &
  &
(photons cm$^{-2}$ s$^{-1}$) &
  &
(erg s$^{-1}$) &
(erg s$^{-1}$) \\

\hline

971110 & 0.0204 & $17.4\pm0.2$ & $0.6\pm0.2$ & $(6.6\pm5.2) \times 10^{51}$ & $(3.8\pm2.1) \times 10^{50}$ \\
990123 & 0.0113 & $16.6\pm0.2$ & 1.6\footnotemark[4] & $(6.6\pm1.0) \times 10^{52}$ & $(7.8\pm2.6) \times 10^{51}$ \\
940526 & 0.0175 & $14.4\pm0.3$ & $0.6\pm0.2$ & $(5.5\pm4.3) \times 10^{51}$ & $(1.8\pm0.7) \times 10^{50}$ \\
980420 & 0.0267 & $3.7\pm0.1$  & $1.1\pm0.3$ & $(7.2\pm4.8) \times 10^{51}$ & $(1.0\pm0.4) \times 10^{51}$ \\
960428 & 0.0628 & $4.2\pm0.1$  & $1.6\pm0.5$ & $(1.9\pm1.4) \times 10^{52}$ & $(2.5\pm1.1) \times 10^{51}$ \\
980105 & 0.0671 & $10.6\pm0.2$ & $1.2\pm0.4$ & $(2.4\pm1.9) \times 10^{52}$ & $(1.2\pm0.7) \times 10^{51}$ \\
980301 & 0.0232 & $0.9\pm0.1$  & $2.4\pm0.8$ & $(9.8\pm5.7) \times 10^{51}$ & $(2.6\pm5.0) \times 10^{51}$ \\
970417a\footnotemark[5]& $\cdots$ & $0.7\pm0.1$  & $0.7\pm0.2$\footnotemark[6] & $(4.3\pm3.1) \times 10^{50}$ & $(0.2\pm1.4) \times 10^{51}$ \\

\end{tabular}
\end{ruledtabular}
\footnotetext[1]{Available at http://www.batse.msfc.nasa.gov/batse/}
\footnotetext[2]{The redshift uncertainties are crudely estimated as
$z/3$ from the variability-luminosity relation.}
\footnotetext[3]{We quote the isotropic luminosities.  These must be 
multiplied by $\Omega/(4\pi)$ to get the true luminosities.}
\footnotetext[4]{Kulkarni et al. (1999)}
\footnotetext[5]{GRB 970417a was too weak in the {\it BATSE} band to apply
the variability-luminosity relation.}
\footnotetext[6]{Totani (2000).}

\end{table}

\clearpage

\begin{table}
\footnotesize
\caption{Internal \& Intergalactic Optical Depths \label{optdeptab}}

\begin{ruledtabular}
\begin{tabular}{cccccc}
   &
\multicolumn{2}{c}{Internal Optical Depth} &   &
\multicolumn{2}{c}{Intergalactic Optical Depth} \\
\cline{2-3} \cline{5-6} \\
 GRB &
 $E_\gamma=100$ GeV &
 1 TeV &   &
 $E_\gamma=100$ GeV &
 1 TeV \\

\hline

971110 & $2.7\pm3.5$ & $2.9\pm3.6$ & & $0.4\pm0.2$ & $9.6\pm4.0$ \\
990123 & $\cdots$\footnotemark[1] & $\cdots$\footnotemark[1] & & 2.8 & 31. \\
940526 & $\cdots$\footnotemark[1] & $\cdots$\footnotemark[1] & & $0.4\pm0.2$ & $9.6\pm4.0$ \\
980420 & $\cdots$\footnotemark[1] & $\cdots$\footnotemark[1] & & $1.3\pm0.8$ & $20.\pm6.$ \\
960428 & $\cdots$\footnotemark[1] & $\cdots$\footnotemark[1] & & $3.1\pm2.0$ & $30.\pm10.$ \\
980105 & $\cdots$\footnotemark[1] & $\cdots$\footnotemark[1] & & $1.6\pm1.2$ & $22.\pm8.$ \\
980301 & $\cdots$\footnotemark[1] & $\cdots$\footnotemark[1] & & $6.8\pm3.4$ & $44.\pm13.$ \\
970417a & $\cdots$\footnotemark[1] & $\cdots$\footnotemark[1] & & $0.5\pm0.3$ & $12.\pm4.$ \\

\end{tabular}
\end{ruledtabular}
\footnotetext[1]{Large uncertainties in the fit parameters made 
constraints on the internal optical depth uninformative in these cases.  
For these bursts, we have made the optimistic assumption 
$\tau_{\gamma\gamma,int}=0.$}

\end{table}

\clearpage

\begin{table}
\footnotesize
\caption{Inferred Properties of Inverse-Compton Model \label{ictab}}

\begin{ruledtabular}
\begin{tabular}{ccccccc}
   &
 $A_\gamma^{IC}$ &
 $E_{cool}^{ob}$ &
 $E_{KN}^{ob}$ &
 $E_{cool}^{KN,ob}$ &
 $E_\gamma^{Tot,{\rm IC}}$ \footnotemark[1] &
 $E_e^{Tot,{\rm IC}}$ \footnotemark[1] \\
 GRB &
 (cm$^{-2}$ s$^{-1}$ GeV$^{-1}$) &
 (GeV) &
 (GeV) &
 (GeV) &
 (erg) &
 (erg) \\

\hline

971110 & $4.\pm80.$ & $(1.\pm13.)\times 10^7$ & $25.\pm7.$ & $75.\pm48.$ & $(2.6\pm8.6)\times 10^{55}$ & $(1.\pm15.)\times 10^{60}$ \\
990123 & $<6.2 \times 10^{-2}$ & 0.04 & 2.8 & 23. & $<3.5 \times 10^{55}$ & $<2.6 \times 10^{57}$ \\
940526 & $<1.2 \times 10^{-2}$ & 14. & 7.3 & 5.1 & $<2.5 \times 10^{54}$ & $<2.5 \times 10^{54}$ \\
980420 & $<4.2 \times 10^{-2}$ & 0.3 & 9.8 & 56. & $<2.4 \times 10^{55}$ & $<1.6 \times 10^{56}$ \\
960428 & $<1.7 \times 10^{-2}$ & 0.004 & 8.2 & 380 & $<2.0 \times 10^{56}$ & $<2.4 \times 10^{59}$ \\
980105 & $<9.6 \times 10^{-2}$ & 0.04 & 17. & 350 & $<3.9 \times 10^{55}$ & $<1.4 \times 10^{57}$ \\
980301 & $<2.1 \times 10^{-1}$ & 0.05 & 6.5 & 73. & $<3.4 \times 10^{56}$ & $<1.0 \times 10^{58}$ \\
970417a& $<1.3 \times 10^{-2}$ & 8.6 & 33. & 63. & $<2.8 \times 10^{54}$ & $<2.8 \times 10^{54}$ \\

\end{tabular}
\end{ruledtabular}
\footnotetext[1]{We quote the isotropic energies.  These entries must be 
multiplied by $\Omega/(4\pi)$ to get the true energies.}

Here we estimate the total energy escaping in the 
gamma-ray component from each burst using the inverse-Compton model.  
We also estimate the total energy in 
the electron component at the source.

\end{table}

\clearpage

\begin{table}
\footnotesize
\caption{Inferred Properties of Proton-Synchrotron Model \label{psynctab}}

\begin{ruledtabular}
\begin{tabular}{ccccc}
   &
 $A_p^{p-sync}$ &
 $E_{\gamma b}^{ob}$ &
 $E_\gamma^{Tot,p-sync}$ \footnotemark[1] &
 $E_p^{Tot,p-sync}$ \footnotemark[1] \\
 GRB &
 (cm$^{-2}$ s$^{-1}$ GeV$^{-1}$) &
 (GeV) &
 (erg) &
 (erg) \\

\hline

971110 & $1.\pm22.$ & $(1.\pm22.) \times 10^{2}$ & $(3.\pm10.) \times 10^{55}$ & $(1.\pm21.) \times 10^{58}$ \\
990123 & $<1.9 \times 10^{-2}$ & $6.2 \times 10^{-5}$ & $<9.6 \times 10^{55}$ & $<9.6 \times 10^{55}$ \\
940526 & $<1.4 \times 10^{-2}$ & $1.3 \times 10^{-2}$ & $<5.3 \times 10^{54}$ & $<5.3 \times 10^{54}$ \\
980420 & $<4.1 \times 10^{-2}$ & $8.9 \times 10^{-4}$ & $<5.4 \times 10^{55}$ & $<5.4 \times 10^{55}$ \\
960428 & $<2.0 \times 10^{-2}$ & $5.8 \times 10^{-5}$ & $<2.7 \times 10^{56}$ & $<2.7 \times 10^{56}$ \\
980105 & $<3.8 \times 10^{-2}$ & $4.0 \times 10^{-4}$ & $<5.6 \times 10^{55}$ & $<5.6 \times 10^{55}$ \\
980301 & $<9.3 \times 10^{-2}$ & $2.5 \times 10^{-4}$ & $<7.0 \times 10^{56}$ & $<7.0 \times 10^{56}$ \\
970417a& $<6.7 \times 10^{-2}$ & $3.0 \times 10^{-2}$ & $<5.9 \times 10^{54}$ & $<5.9 \times 10^{54}$ \\

\end{tabular}
\end{ruledtabular}
\footnotetext[1]{We quote the isotropic energies.  These entries must be 
multiplied by $\Omega/(4\pi)$ to get the true energies.}

Here we estimate the total energy escaping in the 
gamma-ray component from each burst using the proton-synchrotron model.  
We also estimate the total energy in 
the proton component at the source.

\end{table}

\clearpage

\begin{table}
\footnotesize
\caption{Inferred Properties of Photo-Pion Model \label{photoptab}}

\begin{ruledtabular}
\begin{tabular}{ccccc}
   &
 $A_p^\pi$ &
 $E_{\gamma b}^{ob}$ &
 $E_\gamma^{Tot,\pi}$ \footnotemark[1] &
 $E_p^{Tot,\pi}$ \footnotemark[1] \\
 GRB &
 (cm$^{-2}$ s$^{-1}$ GeV$^{-1}$) &
 ($10^5$ GeV) &
 (erg) &
 (erg) \\

\hline

971110 & $(1.\pm13.) \times 10^{11}$ & $13.\pm4.$ & $(4.\pm27.) \times 10^{61}$
 & $(2.\pm27.) \times 10^{69}$ \\
990123 & $<1.3 \times 10^{20}$ & 1.5 & $<1.2 \times 10^{74}$ & $<1.1 \times 10^{78}$ \\
940526 & $<1.8 \times 10^{13}$ & 4.0 & $<2.0 \times 10^{65}$ & $<1.3 \times 10^{70}$ \\
980420 & $<7.8 \times 10^{15}$ & 5.3 & $<2.5 \times 10^{69}$ & $<2.1 \times 10^{73}$ \\
960428 & $<3.9 \times 10^{18}$ & 4.4 & $<5.3 \times 10^{73}$ & $<1.1 \times 10^{77}$ \\
980105 & $<2.3 \times 10^{17}$ & 9.4 & $<1.8 \times 10^{71}$ & $<7.8 \times 10^{74}$ \\
980301 & $<3.1 \times 10^{27}$ & 3.5 & $<3.4 \times 10^{81}$ & $<4.8 \times 10^{85}$ \\
970417a& $<5.8 \times 10^{12}$ & 18. & $<3.2 \times 10^{64}$ & $<1.2 \times 10^{69}$ \\

\end{tabular}
\end{ruledtabular}
\footnotetext[1]{We quote the isotropic energies.  These entries must be 
multiplied by $\Omega/(4\pi)$ to get the true energies.}

Here we estimate the total energy escaping in the 
gamma-ray component from each burst using the photo-pion model.  
We also estimate the total energy in 
the proton component at the source.

\end{table}

\end{document}